\documentclass[11pt,draftcls,onecolumn]{IEEEtran}
\usepackage{amssymb}
\usepackage{cite}
\usepackage{multirow}
\usepackage{tabu}
\usepackage{caption}
\usepackage{lineno}
\usepackage{makecell}
\usepackage{xcolor}
\usepackage{tikz}
\usepackage{threeparttable}
\ifCLASSOPTIONcompsoc
\usepackage[nocompress]{cite}
\else
\usepackage{cite}
\fi

\usepackage[cmex10]{amsmath}
\usepackage{amsthm}
\usepackage{booktabs}
\newtheorem{theorem}{Theorem}
\newtheorem{lemma}[theorem]{Lemma}

\newtheorem{corollary}[theorem]{Corollary}
\theoremstyle{definition}%
\newtheorem{definition}{Definition}
\theoremstyle{remark}%
\newtheorem{remark}{Remark}
\newtheorem{example}{Example}

\newcommand{\fq}{\mathbb{F}_q}

\date{}
\newcommand{\itau}{L}
\renewcommand{\baselinestretch}{1.5}

\title{Weight Distribution of Repeated-Root Cyclic Codes with Prime Power Lengths}

\author{Wei Zhao, Weixian Li, Shenghao Yang, Fang-Wei Fu and Kenneth W. Shum
	\thanks{Wei Zhao is with the School of Mathematics, Foshan University, Guangdong, 528000, China
		(e-mail: zhaowei@fosu.edu.cn).}
	\thanks{Weixian Li is with the School of Mathematics and Statistics, Zhaoqing University, Guangdong, 526061, China  (email: liweixian@zqu.edu.cn).    }
	\thanks{Shenghao Yang and Kenneth W. Shum are with the School of Science and Engineering, The Chinese University of Hong Kong, Shenzhen, Guangdong, 518172, China (e-mail:
		shyang@cuhk.edu.cn, wkshum@cuhk.edu.cn).}
	\thanks{Fang-Wei Fu is with the Chern Institute of Mathematics and LPMC, Nankai University, Tianjin 300071, China (email:  fwfu@nankai.edu.cn).}
}

\begin{document}

\maketitle

\begin{abstract}
Determining the weight distribution of a linear code is a classical and fundamental topic in coding theory that has been extensively investigated. Repeated-root cyclic codes, which form a significant subclass of error-correcting codes, have found broad applications in quantum error-correcting codes, symbol-pair codes, and storage codes. Through polynomial derivation, we derive the monomial equivalent codes for these repeated-root cyclic codes with prime power lengths. Given that monomial equivalent codes exhibit identical weight distributions, we transform the computation of the weight distribution of these repeated-root cyclic codes into the computation of the weight distribution of their monomial equivalent codes. Leveraging the classical results on the weight distribution of MDS codes, we explicitly determine the weight distribution of these repeated-root cyclic codes. Moreover, we apply the weight distribution formula to construct a class of $p$-weight cyclic codes for any prime $p$.
\end{abstract}

\section{Introduction}
Let $\mathcal{C}$ be a linear code of length $n$ over a finite field $\mathbb{F}_q$. The Hamming weight of a codeword is the number of nonzero coordinates of the codeword. For a positive integer $i$ such that $1\leq i\leq n$, let $A_i$ denote the number of nonzero codewords with Hamming weight $i$ in $\mathcal{C}$. The sequence $(1,A_1, \ldots,A_n)$ is called the weight distribution of $\mathcal{C}$. The polynomial $1+A_1z+A_2z^2+\cdots+A_nz^n$ is referred to as the weight enumerator of $\mathcal{C}$. The weight distribution of a code is a fundamental property that not only gives the error correcting ability of the code, but also enables the computation of the error probability for both error detection and correction \cite{klove2012error}.

Cyclic codes, a class of linear block codes whose set of codewords is closed under cyclic shifting, enjoys a prominent place in the theory of error-correcting codes. The well-known codes, including Reed-Solomon, BCH, Golay, Reed-Muller, and binary Hamming codes, are either cyclic codes or derived from cyclic codes \cite{macwilliams1977theory}. Owing to their algebraic structure, cyclic codes can be efficiently encoded using shift registers, which accounts for their preferred role in engineering applications. Determining the weight distribution of cyclic codes is typically a challenging problem that has yet to be fully resolved, and remains open for the majority of cyclic codes.

The weight distribution of a cyclic code is basically connected to the value distribution of a certain exponential sum. To date, through the application of Gauss sums, Gauss periods, quadratic forms, Niho exponents and elliptic curves, researchers in coding theory have successfully determined the weight distributions of several cyclic codes with specific zero set. For the duals of primitive cyclic codes with few quadratic type zeros, their weight distributions have been intensively studied, for example see \cite{ding2013five,li2014weight,liu2015class,zheng2014weight,zheng2015weight}. For the duals of primitive cyclic codes with few non-quadratic type zeros, readers can refer to \cite{lis2014weight,li2010class,mcguire2004three,vega2012weight}. For the duals of non-primitive cyclic codes with few zeros, their weight distributions have been discussed in \cite{yuan2006weight,li2013weight,wang2012weight}. Despite the significant progress made in recent years, the weight distributions of most cyclic codes remain unknown.
In particular, there is a critical need for the development of novel mathematical tools and methodologies to accurately determine the weight distributions of a broader spectrum of cyclic codes.

Repeated-root cyclic codes, as an important subclass of error-correcting codes, were initially studied in 1967 by Berman \cite{berman1967semisimple}. During the 1990s, Castagnoli \textit{et al.} \cite{castagnoli1991repeated} and van Lint \cite{van1991repeated} conducted a comprehensive investigations into repeated-root cyclic codes. A formula of minimum distance of repeated-root cyclic codes was established in \cite{castagnoli1991repeated}. 
van Lint in \cite{van1991repeated} showed that a binary cyclic code of length $2n$ ($n$ odd) can be obtained from two cyclic codes of length $n$ by the $|u|u + v|$ construction. Repeated-root cyclic codes are optimal in a few cases, which has motivated researchers to further study
this class of codes, for example \cite{dinh2005negacyclic,dinh2007complete,dinh2008linear,liu2017some,Sobhani2016}.
Let $p$ be a prime, repeated-root cyclic codes of length $p^s$ over a finite field $\mathbb{F}_{p^m}$ have a very interesting
structure. Dinh \cite{dinh2010constacyclic} showed that the cyclic codes of length $p^s$ are linearly ordered under set-theoretic inclusion as the ideals of the ring $\mathbb{F}_{p^m}[x]/\langle x^{p^s}-1\rangle$. As an application, Dinh \cite{dinh2010constacyclic} obtained the exact values of minimum Hamming distances of such codes. These results have been widely used in constructing maximum distance separable symbol-pair codes \cite{dinh2017symbol,dinh2019mds,ma2022mds}.
Recent research has expanded the investigation of repeated-root cyclic codes into diverse applications and generalizations. Luo and Ma constructed non-binary quantum synchronizable codes from repeated-root cyclic codes, demonstrating that these codes offer superior error-correcting capabilities compared to BCH codes in specific scenarios~\cite{luo2018non}. Additionally,  repeated-root cyclic codes and related codes have been applied in storage systems, where they demonstrate favorable storage properties \cite{yu2024,hou2022}. 

The study of the weight distribution of repeated-root cyclic codes presents a significant challenge. By extending the $|u|u + v|$ construction \cite{van1991repeated} to a squaring construction, Nedeloaia \cite{nedeloaia2003weight} reduced the problem of determining the weight distribution of a repeated-root cyclic code with even code length $n$ and dimension $2k_A + k_B$ to that of computing the weight distributions of $2^{k_B}$ linear codes of length $\frac{n}{2}$ and dimension $k_A$. Using the Magma software, the author claimed the ability to compute the weight distribution of repeated-root cyclic codes of length up to 110. Similarly, the work in \cite{bhullar2010repeated} applied an analogous method to simplify the computation of weight distributions for repeated-root cyclic codes based on a two-level squaring construction. However, neither Nedeloaia \cite{nedeloaia2003weight} nor Bhullar \textit{et al.} \cite{bhullar2010repeated} provided explicit formulas for calculating the weight distribution of repeated-root cyclic codes. Dinh \cite{dinh2008linear} derived an explicit weight distribution formula for certain specific repeated-root cyclic codes with prime power lengths. Nevertheless, as of now, only limited results have been reported on the weight distribution of repeated-root cyclic codes in general. In this paper, we conduct a systematic investigation into the weight distribution of all repeated-root cyclic codes with prime power lengths. The weight distribution formula presented in \cite{dinh2008linear} can be derived as a special case of our general results.

A linear code is referred to as a $t$-weight linear code if the number of nonzero $A_i$ in the sequence $(A_1,A_2, \ldots,A_n)$ is equal to $t$.
Linear codes with few weights have many applications in various fields, including secret sharing \cite{AndersonDHK98, CarletDY05}, strongly regular graphs \cite{ShiHSQWS21}, association schemes \cite{calderbank1984three} and authentication codes \cite{DingHKW07}. Significant progress has been made in the construction of two-weight, three-weight, and four-weight linear codes. For instance, the relevant contributions can be found in \cite{DingD15,TangLQZH16,ZhouLFH16,ShiLH24} and the references therein. However, there has been limited research progress regarding the construction of $t$-weight linear codes for any positive integer $t$. By using the weight distribution formula of the repeated-root cyclic codes, we introduce a class of $p$-weight cyclic codes for any prime $p$.
 
In this paper, we investigate the computation of weight distributions of repeated-root cyclic codes with prime power lengths. Our main contributions are summarized as follows:
\begin{itemize}
	\item[1.] We present a new structural characterization of repeated-root cyclic codes with prime power lengths, see Theorem \ref{lem3.1} and Theorem \ref{ciProperty}. In comparison to the existing structural results for such codes (see, e.g., \cite{castagnoli1991repeated, van1991repeated, dinh2005negacyclic, dinh2007complete}), our results provide a more detailed description of the composition of each individual codeword. This advancement contributes to a deeper understanding of these codes and help to solve the weight distribution problem of these codes.
	\item[2.] We present the weight distribution of all repeated-root cyclic codes with prime power lengths, as stated in Theorem \ref{th1}, Theorem \ref{th3}, Theorem \ref{rest}, Theorem \ref{dual} and Theorem \ref{th2}. The results concerning the weight distribution of repeated-root cyclic codes provided in \cite{dinh2008linear} are special cases of our findings, as detailed in Remark \ref{rem1}, Remark \ref{rem2}, Corollary \ref{cor1} and Corollary \ref{cor2}. Compared to the results proposed in \cite{nedeloaia2003weight}, which reduces the computation of the weight distribution of cyclic codes to computations of the weight distributions of several (specifically, a power of 2 number of) linear codes with half the length and lower dimensions, we present an explicit formula for the weight distribution of certain classes of repeated-root cyclic codes with prime power lengths, see Theorem \ref{th1}, Theorem \ref{th3}, Theorem \ref{rest} and Theorem \ref{dual}. For the remaining cases, the computation is significantly simplified to determining a single, smaller set $I$, as established in Theorem \ref{th2}. We also introduce a class of $p$-weight cyclic codes for any prime $p$ in Theorem \ref{constant-weight}.
\end{itemize}

The rest of this paper is organized as follows. In Section II, we review some fundamental results
on repeated-root cyclic codes. In Section III, we investigate the monomial equivalent codes of repeated-root cyclic codes with prime power lengths. The weight distribution of these codes is divided into two cases based on their structural properties, which are presented in Sections IV and V, respectively.
In Section VI, we conclude this paper.

\section{Preliminaries}\label{sec:preliminaries}
Denote the $n$-dimensional vector space over $\mathbb{F}_{q}$ as $\mathbb{F}_{q}^{n}$. A {\em linear code} of length $n$ over $\mathbb{F}_{q}$ is a subspace of $\mathbb{F}_{q}^{n}$. A vector of a linear code is referred to as a {\em codeword}.
Define the cyclic shift operator $\psi$ on $\mathbb{F}_{q}^{n}$ by
$$
\psi (v_{0}, v_{1}, \ldots, v_{n-1}):= (v_{n-1}, v_{0}, \ldots, v_{n-2})
$$
for a vector $\boldsymbol{v}=(v_{0}, v_{1}, \ldots, v_{n-1})$ in $\mathbb{F}_{q}^{n}$.
Let $\langle x^n-1 \rangle$ denote the ideal of $\fq[x]$ generated by the polynomial $x^n-1$.
For the residue ring $\mathbb{F}_{q}[x]/\left\langle x^{n}-1\right\rangle$, we adopt the polynomial of the smallest degree to uniquely represent the equivalence class it belongs to. Specifically, the class $v_{0}+v_{1} x+\cdots+v_{n-1} x^{n-1}+\left\langle x^{n}-1\right\rangle$ is represented as $v_{0}+v_{1} x+\cdots+v_{n-1} x^{n-1}$.
Consequently, there exists a one-to-one correspondence between the elements in the residue ring and the vectors in $\mathbb{F}_q^n$. More precisely,
we represent a vector $\boldsymbol{v}= (v_{0}, v_{1}, \ldots, v_{n-1})$ in $\mathbb{F}_q^n$ as a polynomial
$$
v(x):= v_{0}+v_{1} x+\cdots+v_{n-1} x^{n-1}\in \mathbb{F}_{q}[x]/\left\langle x^{n}-1\right\rangle.
$$
The polynomial $v(x)$ is referred to as the polynomial representation of the vector $\boldsymbol{v}$.
Under this framework, the multiplication operation $x v(x)$ in ${\mathbb{F}_{q}[x]}/{\langle x^{n}-1\rangle}$ corresponds to performing a cyclic shift on the vector~$\boldsymbol{v}$.
A linear code $\mathcal{C}$ is said to be \textit{cyclic} if it is closed under the cyclic shift, i.e., $\psi(\textbf{c})\in \mathcal{C}$ for all $\textbf{c}\in \mathcal{C}$.
Hence, a cyclic code $\mathcal{C}$ of length $n$ over $\mathbb{F}_{q}$ corresponds to an ideal of the quotient ring
${\mathbb{F}_{q}[x]}/{\left\langle x^{n}-1\right\rangle}$.
Algebraically, a cyclic code $\mathcal{C}$ of length $n$ can be represented as an ideal:
$$ \mathcal{C} = \langle g(x)\rangle =  \left\{f(x)g(x)\in {\mathbb{F}_{q}[x]}/{\langle x^n-1\rangle}|\text{ for all }f(x)\in {\mathbb{F}_{q}[x]}/{\langle x^n-1\rangle}\right\},$$
where $g(x)$ is monic and a divisor of $x^n-1$. The unique polynomial $g(x)$ is called the {\em generator polynomial} of the cyclic code~$\mathcal{C}$.
The dimension of $\mathcal{C}$ is $n-\text{deg}(g(x))$, where $\text{deg}(g(x))$ denotes the degree of $g(x)$.
When the code length $n$ and the field size $q$ are not relatively prime, the generator polynomial $g(x)$ can have repeated roots.

In this paper, we are interested in a family of cyclic codes with prime power length~$n=p^s$ over finite field $\mathbb{F}_q$, where $p$ is the characteristic of $\mathbb{F}_q$ and $s$ is a positive integer. These cyclic codes are denoted as $\mathcal{C}_{i,s} = \langle x^i - 1 \rangle \subseteq {\mathbb{F}_{q}[x]}/{\langle x^{p^s}-1\rangle} $, where $0 \leq i \leq p^s$. When the value of the integer $s$ is clear from context, the notation may be simplified to $\mathcal{C}_{i}$.

In the specific cases where $i=0$ and $i=p^s$, we have the trivial codes $\mathcal{C}_{0}=\left\langle 1 \right \rangle$ and $\mathcal{C}_{p^s}=\left\langle 0 \right \rangle$, which correspond to minimum distances $1$ and $0$, respectively. Furthermore, the weight distribution of $\mathcal{C}_{0}$ is $A_0=1$ and $A_m=\binom{p^s}{m}(q-1)^m$ for $1\leq m\leq p^s$, and the weight distribution of $\mathcal{C}_{p^s}$ is $A_0=1$ and $A_m=0$ for $1\leq m\leq p^s$. To facilitate rigorous treatment of these cases and enhance the clarity of forthcoming proofs, we formally introduce some mathematical notations and conventions in Table \ref{notations}.
\begin{table}[h]
	\centering
	\caption{Mathematical Notations and Conventions}
	\label{notations}
\begin{center}
	\begin{tabular}{r|l}
		\hline
		$s$  & a positive integer;\\
		$p$  & a prime number;\\
		$q$  & a power of $p$;\\
		$i$  & an integer within the interval $[1,p^s-1]$;\\
		$\mathcal{C}_{i,s}$  &  the cyclic code $\mathcal{C}_{i,s} =\langle (x-1)^i\rangle\subseteq {\mathbb{F}_{q}[x]}/{\langle x^{p^s}-1\rangle}$; \\
		$\mathcal{C}_i$  & the cyclic code $\mathcal{C}_{i,s} $ if $s$ is clear from context; \\
		$\hat{\mathcal{C}}_i$  & the cyclic code $\mathcal{C}_{i,1}$;\\
		$t$  & an integer within the range $[0,s-1]$;\\
		$\tau$  & an integer within the range $[1,p-1]$;\\
		$\itau(t,\tau)$  & an integer equal to $p^{s}-p^{s-t}+\tau p^{s-t-1}$;\\
		$S_{t,\tau}$  & a set of positive integers $\{i\in \mathbb{N} |\, \itau(t,\tau-1)< i < \itau(t,\tau)\}$;\\
		$\binom{a}{b}$ & the binomial coefficient and it is defined to be zero when $a<b$;\\
		$\sum\limits_{x\in T}f(x)$ & the summation evaluates to zero if the set $T$ is empty.\\
		\hline
	\end{tabular}
\end{center}
\end{table}

In the rest of this paper, we aim to determine the weight distribution of $\mathcal{C}_{i}$ for $1\leq i\leq p^s-1$.
Firstly, we recall the result of minimum distance of these codes.
\begin{lemma}[\cite{dinh2008linear,Massey1973}] \label{dis}\label{D08}
For $i=1,2,\ldots, p^{s}-1$, the cyclic code $\mathcal{C}_{i,s}$ has the minimum distance given by
\[ d_i = (\tau+1)p^t,
\]
where $\tau$ and $t$ are unique integers satisfying $1 \le \tau \le p-1$, $0 \le t \le s-1$, and
$$p^{s}-p^{s-t}+(\tau-1) p^{s-t-1} < i \le p^{s}-p^{s-t}+\tau p^{s-t-1}.$$
\end{lemma}
\begin{example}
	Consider the cyclic codes of length $n=25$ over $\mathbb{F}_5$ generated by polynomials $(x-1)^i$, for $i=1,\ldots, 24$. The minimum distance $d_{\min}$ of $\mathcal{C}_i$ are given in the following table:
	\begin{center}
		\begin{tabular}{|c||c|c|c|c|c|c|c|c|} \hline
			$i$ &  $1,\ldots, 5$ & $6,\ldots,10$ & $11,\ldots, 15$ & $16,\ldots, 20$ & $21$ & $22$ & $23$ & $24$ \\ \hline
			$d_{\min}$ & $2$ & $3$ & $4$ & $5$ & $10$ & $15$ & $20$ & $25$  \\ \hline
		\end{tabular}
	\end{center}
\end{example}

By specializing Lemma~\ref{D08} to the case where $s=1$, we can infer an MDS code. We formalize this conclusion in the following lemma.
\begin{lemma}
\label{lemma:MDS_length_p}
For $1\leq i \leq p-1$, the repeated-root cyclic code $\hat{\mathcal{C}}_i=\langle (x-1)^i \rangle\subseteq \mathbb{F}_q[x]/\langle x^{p}-1\rangle$ is an MDS code with parameters $[p, p-i, i+1]$.
\end{lemma}

\begin{proof}
Verify that the code length of $\hat{\mathcal{C}}_i$ is $p$. Given that the degree of the generator polynomial of $\hat{\mathcal{C}}_i$ is $i$,  the dimension is thereby $p-i$. From Lemma \ref{dis}, we deduce that $t=0$ and $\tau=i$; consequently, the minimum distance $d_i=i+1$. Therefore, $\hat{\mathcal{C}}_i$ satisfies the parameters of an MDS code.
\end{proof}

The weight distribution of MDS codes has been established in \cite{s1}. Given that $\hat{\mathcal{C}}_i$ is an MDS code, its weight distribution can be explicitly determined.

\begin{theorem}\label{lem1}
	For $1\leq i\leq p-1$, the weight distribution of $\hat{\mathcal{C}}_i$ is given by $A_0=1$, $A_m=0$ for $1 \leq m \leq i$, and
	$$
	A_m=\binom{p}{m} \sum_{j=0}^{m-i-1}(-1)^j\binom{m}{j}\left(q^{m-i-j}-1\right),
	$$
	for $i+1 \leq m \leq p$.
\end{theorem}
\begin{proof}
     The result can be directly obtained by applying the parameters of $\hat{\mathcal{C}}_i$, as presented in Lemma \ref{lemma:MDS_length_p}, to the weight distribution formula for MDS codes given in \cite{s1} .
\end{proof}

The aforementioned theorem determines the weight distribution of repeated-root cyclic codes for length~$p$. For lengths $p^s$ where $s\geq 2$, the associated cases demonstrate increased complexity and require more detailed investigation.

\section{Monomially Equivalent Codes}
The method for analyzing the weight distribution of cyclic codes presented in this paper is to transfer the problem into analyzing the weight distribution of their monomially equivalent codes, which proves to be more tractable. Firstly, we investigate the monomially equivalent codes of $\mathcal{C}_i=\langle (x-1)^i\rangle\subseteq\mathbb{F}_q[x]/\langle x^{p^s}-1\rangle$. Recall that a monomial matrix is defined as a square matrix with precisely one nonzero entry in each row and each column. It is worth noting that a monomial matrix $M$ can be decomposed into $M = DP$, where $D$ represents a diagonal matrix with nonzero diagonal entries, and $P$ denotes a permutation matrix.
The definition of the monomially equivalent relation is presented as follows.

\begin{definition}
	Two linear codes, $\mathcal{D}_1$ and $\mathcal{D}_2$, are said to be {\em monomially equivalent} if there exists a monomial matrix $M$ such that
	the product $G_1M$ constitutes a generator matrix for $\mathcal{D}_2$, where $G_1$ denotes the generator matrix of $\mathcal{D}_1$.
\end{definition}

It is straightforward that the monomially equivalent codes possess the same weight distribution. For a linear code $\mathcal{C}$ over $\mathbb{F}_{q}$, we define the repetition and the direct sum methods for constructing new codes from $\mathcal{C}$ as follows. Let $m$ be a positive integer.
The $m$-th repetition code $\mathcal{C}^m$ of $\mathcal{C}$ is defined by
$$\mathcal{C}^m:=\{(\underbrace{\mathbf{c},\mathbf{c},\cdots,\mathbf{c}}_{m \text{ times}}) | \, \mathbf{c} \in \mathcal{C} \}.$$
The $m$-th direct sum code $\mathcal{C}^{\oplus m}$ of $\mathcal{C}$ is defined by
$$\mathcal{C}^{\oplus m}:=\{(\mathbf{c}_1,\mathbf{c}_2,\cdots,\mathbf{c}_m) | \, \mathbf{c}_i \in \mathcal{C} \text{ for } i = 1,2,\cdots,m \}.$$
It is easy to see that $(\mathcal{C}^m)^{\oplus n}$ and $(\mathcal{C}^{\oplus n})^{m}$ are monomially equivalent.
For an $n\times n$ matrix $A=\left(a_{i,j}\right )$ and an $m\times m$ matrix $B$, the tensor product of $A$ and $B$ is the $nm\times nm$ matrix
\[A\otimes B:=\left[ \begin{matrix}
	{{a}_{1,1}}B & {{a}_{1,2}}B & \cdots  & {{a}_{1,n}}B  \\
	{{a}_{2,1}}B & {{a}_{2,2}}B & \cdots  & {{a}_{2,n}}B  \\
	\vdots  & \vdots  & \vdots  & \vdots   \\
	{{a}_{n,1}}B & {{a}_{n,2}}B & \cdots  & {{a}_{n,n}}B  \\
\end{matrix} \right].
\]

\begin{theorem}\label{d}
For $0< i^{'}\leq p^{s-t-1}$, $\mathcal{C}_{\itau(t,\tau-1)+i^{'},\,s}$ is monomially
		equivalent to $\mathcal{C}^{p^t}_{(\tau-1) p^{s-t-1}+i^{'},\,s-t}$.
\end{theorem}
\begin{proof}
Denote $\mathcal{C}$ and $\mathcal{D}$ as the cyclic codes $\mathcal{C}_{\itau(t,\tau-1)+i^{'},\,s}$ and  $\mathcal{C}^{p^t}_{(\tau-1) p^{s-t-1}+i^{'},\,s-t}$, respectively. We need to prove that $\mathcal{C}$ is monomially
equivalent to $\mathcal{D}^{p^t}$.
Let $c(x)=(x-1)^{\itau(t,\tau-1)+i^{'}}f(x)$ be a codeword in~$\mathcal{C}$, where $f(x)\in \mathbb{F}_q[x]$ and
the degree of $f(x)$ is less than $(p-\tau+1)p^{s-t-1}-i^{'}$. Denote
$h(x)=(x-1)^{(\tau-1) p^{s-t-1}+i^{'}}f(x)$. Then, the degree of $h(x)$ is
less than $p^{s-t}$, and $c(x)$ is transformed into:
\begin{align} 
	c(x) & = (x-1)^{p^s - p^{s-t}}h(x) \nonumber \\
	&=(x^{p^{s-t}}-1)^{p^t - 1}h(x) \nonumber 
	\\
	&=\left[\sum_{j=0}^{p^{t}-1} \binom{p^{t} -1}{j} (-1)^{(p^{t}-1-j)} x^{j p^{s-t}}\right] h(x) \label{3} \\
	&= \sum_{j=0}^{p^{t}-1}\left[ \binom{p^{t} -1}{j} (-1)^{(p^{t}-1-j)} h(x)\right] x^{j p^{s-t}},  \label{eq:4}
\end{align}
where the final equality holds because the degree of $ h(x) $ is strictly less than $ p^{s-t} $.  
According to Lucas's theorem, $p$ cannot divide $\binom{p^{t} -1}{j}$. Hence $\binom{p^{t} -1}{j}(-1)^{(p^{t}-1-j)}$
in \eqref{3} is a nonzero element in $\mathbb{F}_q$ for all integers $j$ with $0\leq j\leq p^t-1$. By observing \eqref{3}, for each $0\leq j\leq p^t-1$, the degree of any monomial in the expansion of $\binom{p^{t} -1}{j} (-1)^{(p^{t}-1-j)} x^{j p^{s-t}} h(x) $ is in the interval $[jp^{s-t},(j+1)p^{s-t})$. 
Consider the expression $\binom{p^{t} - 1}{j} (-1)^{p^{t} - 1 - j} h(x)$ in Equation \eqref{eq:4}. As $c(x)$ ranges over all codewords of $\mathcal{C}$, this expression spans all codewords of $\mathcal{D}$ for each $j$ satisfying $0 \leq j \leq p^t - 1$. This holds because the set  
$$
\left\{ \binom{p^{t} - 1}{j} (-1)^{p^{t} - 1 - j} (x - 1)^{\tau p^{s - t - 1} + i'} f(x) \,\middle|\, f(x) \in \mathbb{F}_q[x] \text{ and } \deg(f(x)) < (p - \tau + 1)p^{s - t - 1} - i' \right\}
$$  
is exactly equal to $\mathcal{D}$ for all $j$ in the range $0 \leq j \leq p^t - 1$.
Therefore, by \eqref{eq:4}, each codeword of $\mathcal{C}$ can be represented by $p^t$ codewords of $\mathcal{D}$.
It can be further demonstrated that $\mathcal{C}$ is monomially
equivalent to $\mathcal{D}^{p^t}$, as detailed below.
Let
\begin{equation*}
	D=\mathrm{diag}\left(\binom{p^{t} -1}{j} (-1)^{(p^{t}-1-j)},j=0,1,\ldots,p^t-1\right)\otimes I_{p^{s-t}},
\end{equation*}
where $\mathrm{diag}(x_1,\ldots,x_n)$ is an $n\times n$ diagonal
matrix with diagonal entries $x_1,\ldots,x_n$, and $I_{p^{s-t}}$ is
the identity matrix of size $p^{s-t}$. Let $G_\mathcal{D}$ denote a
generator matrix of $\mathcal{D}$. Then,
$1_{p^t} \otimes G_\mathcal{D}$ serves as a generator matrix of
$\mathcal{D}^{p^t}$, where $1_{p^t}$ denotes the all-one vector of
length $p^t$ over $\fq$. According to Equation \eqref{eq:4}, $(1_{p^t} \otimes G_\mathcal{D}) D$
constitutes a generator matrix of $\mathcal{C}$, thereby establishing that $\mathcal{C}$ and $\mathcal{D}^{p^t}$ are monomially equivalent.
\end{proof}

\begin{corollary}\cite[Theorem 6.6]{dinh2008linear}
	Let $(A_0(i,s),\ldots,A_{p^s}(i,s))$ be the weight distribution of $\mathcal{C}_{i,\,s}$. Then for $0< i^{'}\leq p^{s-t-1}$,
	the weight distribution of $\mathcal{C}_{\itau(t,\tau-1)+i^{'},\,s}$ is
	$$A_{j}(i,s)=\begin{cases}
		A_{z}((\tau-1) p^{s-t-1}+i^{'},s-t), & \text{if } j=p^{t}z, \text{ for } 0\leq z\leq p^{s-t},\\
		0, & \text{otherwise}.
	\end{cases}$$
\end{corollary}
\begin{proof}
	The conclusion can be directly derived from Theorem \ref{d}.
\end{proof}
\begin{remark}
	The former corollary can also be derived from the approach described in \cite[Theorem 6.6]{dinh2008linear}, which leverages the isomorphism between cyclic and negacyclic codes along with the known weight distribution of negacyclic codes. Due to this isomorphism, the weight distribution results presented in this paper for repeated-root cyclic codes—specifically those given in Theorems \ref{th1}, \ref{th3}, \ref{rest}, \ref{dual}, and \ref{th2} in the following sections—can be directly extended to the corresponding repeated-root negacyclic codes.
\end{remark}

Guided by Theorem~\ref{d}, for lengths $p^s$ where $s\geq 1$, we analyze the weight distribution of $\mathcal{C}_i$ by considering two distinct cases based on the parameter $i\in [1,p^s-1]$:
\begin{enumerate}
	\item[] Case I. $i=\itau(t,\tau)$, where $t\in [0,s-1]$ and $\tau\in [1,p-1]$;
	\item[] Case II. $i\in S_{t,\tau}$, where $t\in [0,s-1]$ and $\tau\in [1,p-1]$.
\end{enumerate}

\section{Weight Distribution for Case I and Its Dual Codes}
We begin by conducting a further investigation into the structural properties of $\mathcal{C}_{\itau(t,\tau)}$, which offers a clearer understanding than the result presented in Theorem~\ref{d}.
\begin{theorem}\label{lem3.1}
For integers $1\leq \tau\leq p-1$ and $0\leq t\leq s-1$, $\mathcal{C}_{\itau(t,\tau)}$ is monomially
equivalent to both $({\hat{\mathcal{C}}}_\tau^{\oplus p^{s-t-1}})^{p^t}$ and $({\hat{\mathcal{C}}}_\tau^{p^t})^{\oplus p^{s-t-1}}$.
\end{theorem}

\begin{proof}
According to Theorem \ref{d}, $\mathcal{C}_{L(t,\tau)}$ is monomially
equivalent to $\mathcal{D}^{p^t}$,
where $$\mathcal{D} = \left\{(x-1)^{\tau p^{s-t-1}}f(x)\,\middle|\, f(x)\in \mathbb{F}_q[x],\, \deg(f)< (p-\tau)p^{s-t-1}\right\}.$$
To further decompose the code $\mathcal{D}$, we rewrite $f(x)$ as follows.
Let $\gamma=p^{s-t-1}$, and let $f_i$ represent coefficients of $f(x)$ such that
$f(x)=\sum_{i=0}^{(p-\tau)\gamma-1}f_{i} x^{i}$. For $0 \leq j \leq \gamma-1$, denote
\begin{equation}\label{fj}
	F_j(x)=\sum_{i=0}^{p-\tau-1}f_{i\gamma+j}x^i.
\end{equation}
Consequently, we obtain the following expression:
\begin{align*}
	f(x) & =\sum_{i=0}^{(p-\tau)\gamma-1}f_{i} x^{i} \\
	&=\sum_{j=0}^{\gamma-1}\left(\sum_{i=0}^{p-\tau-1}f_{i\gamma+j}x^{i\gamma}\right)x^j\\
	&=\sum_{j=0}^{\gamma-1}F_j(x^\gamma)x^j.
\end{align*}
Further, the codeword $h(x)=(x-1)^{\tau p^{s-t-1}}f(x)\in\mathcal{D}$ can be formally transformed into
\begin{align} \label{eq:5}
	h(x)  = \sum_{j=0}^{\gamma-1}\left( (x^{\gamma}-1)^{\tau}F_j(x^{\gamma}) \right)x^j.
\end{align}
For $0 \le j \le \gamma-1$, let
$N_{j}=\{i \gamma+j|\,0\leq i\leq p-1\}$. Let $\mathbf{h} \in \mathbb{F}_q^{p^{s-t}}$ represent the vector corresponding to the polynomial $h(x)$, and let ${\mathbf{h}|}_{N_{j}}$ denote the subvector of $\mathbf{h}$ indexed by the elements of $N_j$. For each~$j\in\{0,1,\ldots,\gamma-1\}$, using Equation (\ref{eq:5}), the polynomial representation of ${\mathbf{h}|}_{N_{j}}$ is given by
$(x-1)^\tau F_j(x)$. 
According to \eqref{fj}, if $f(x)$ ranges over all polynomials of degree less than $(p - \tau)p^{s - t - 1}$, then each $F_j(x)$, for all $0 \leq j \leq p^{s - t - 1} - 1$, ranges over all polynomials of degree less than $p - \tau$.
Consequently, ${\mathcal{D}|}_{N_{j}}$ represents a cyclic code of length $p$ over $\mathbb{F}_{q}$ with generator polynomial $(x-1)^\tau$, i.e., ${\mathcal{D}|}_{N_{j}} = \hat{\mathcal{C}}_\tau$.
We conclude that
$\mathcal{D}$ is monomially equivalent to
$\hat{\mathcal{C}}_\tau^{\oplus p^{s-t-1}}$, with the corresponding
monomial matrix
$P=(\delta_{\alpha \beta})_{p^{s-t} \times p^{s-t}}$, where
$$\delta_{\alpha \beta}=\left\{
\begin{array}{ll}
	1, & \text{ if } \alpha= j p +\lambda \text{ and } \beta = \lambda p^{s-t-1} +j \text{ for } 0 \le \lambda \le p-1 \text{ and } 0 \le j \le p^{s-t-1} -1, \\
	0, & \text{otherwise.}
\end{array}\right.$$
Let $G_{\hat{\mathcal{C}}_\tau}$ denote a generator matrix of $\hat{\mathcal{C}}_\tau$. Then, by combining \eqref{eq:5} and ${\mathcal{D}|}_{N_{j}}=\hat{\mathcal{C}}_\tau$, it follows that $(I_{p^{s-t-1}} \otimes G_{\hat{\mathcal{C}}_\tau} ) P$ serves as a generator matrix of $\mathcal{D}$.
Consequently, $\mathcal{C}_{\itau(t,\tau)}$ is monomially equivalent to $({\hat{\mathcal{C}}_\tau}^{\oplus p^{s-t-1}})^{p^t}$, where the monomial equivalence is established by the matrix
$M=(I_{p^{t}} \otimes P) D$. Furthermore, the matrix $G=\left(1_{p^t}\otimes (I_{p^{s-t-1}} \otimes G_{\hat{\mathcal{C}}_\tau} )\right) (I_{p^{t}} \otimes P) D$ serves as a generator matrix for $\mathcal{C}_{\itau(t,\tau)}$.
Additionally, since $({\hat{\mathcal C}}_\tau^{\oplus p^{s-t-1}})^{p^t}$ is monomially equivalent to $({\hat{\mathcal C}}_\tau^{p^t})^{\oplus p^{s-t-1}}$,
it follows that $\mathcal{C}_{\itau(t,\tau)}$ is monomially equivalent to $({\hat{\mathcal{C}}_\tau}^{p^t})^{\oplus p^{s-t-1}}$.
\end{proof}

Based on Theorem \ref{lem3.1}, it can be inferred that the cyclic code $\mathcal{C}_{\itau(t,\tau)}$ is monomially equivalent to a specific matrix-product code. To begin with, we formally introduce the concept of matrix-product codes. Let $m$, $n$ denote positive integers with $m \le n$. Let $A=(a_{i,j})$ represent an $m \times n$ matrix over the finite field $\mathbb{F}_{q}$. Let $\mathbf{z}_1$,\ldots,$\mathbf{z}_m$ denote $m$ row vectors of length $l$ in $\mathbb{F}_q^{l}$. A multiplication operation~$\odot$ is defined as follows:
\[(\mathbf{z}_1,\mathbf{z}_2,\ldots,\mathbf{z}_m)\odot A:=\left( \sum\limits_{i=1}^{m}{{{\mathbf{z}}_{i}}{{a}_{i,1}}},\ldots ,\sum\limits_{i=1}^{m}{{{\mathbf{z}}_{i}}{{a}_{i,n}}} \right)\in \mathbb{F}_q^{ln}. \]
Let $\mathcal{D}_1,\ldots, \mathcal{D}_m$ be $m$ linear codes of length $n$ over $\mathbb{F}_q$. The matrix-product code, as introduced in \cite{blackmore2001product}, takes the form
$$(\mathcal{D}_1,\ldots,\mathcal{D}_m) \odot A := \left\{ (\mathbf{d}_1,\mathbf{d}_2,\ldots,\mathbf{d}_m)\odot A \bigg{|} \,\mathbf{d}_1 \in \mathcal{D}_1 ,\ldots , \mathbf{d}_m \in \mathcal{D}_m \right\}.$$
In the subsequent discussion, we intend to demonstrate that the cyclic code $\mathcal{C}_{\itau(t,\tau)}$ is monomially equivalent to a specific matrix-product code.
\begin{lemma}\label{mat}
The code $\mathcal{C}_{\itau(t,\tau)}$ is monomially equivalent to 	$$(\underbrace{\hat{\mathcal{C}}_\tau,\ldots,\hat{\mathcal{C}}_\tau}_{p^{s-t-1} \text{ times}}) \odot (I_{p^{s-t-1}} \otimes 1_{p^t})$$
and
$$ (\underbrace{\hat{\mathcal{C}}_\tau,\ldots,\hat{\mathcal{C}}_\tau}_{p^{s-t-1} \text{ times}}) \odot (1_{p^t} \otimes I_{p^{s-t-1}}) .$$
\end{lemma}
\begin{proof}
We observe that the repetition code $\mathcal{D}^m=\mathcal{D}\odot 1_{m}$ and the direct sum code $\mathcal{D}^{\oplus m}=(\underbrace{\mathcal{D},\ldots,\mathcal{D}}_{m \text{ times}}) \odot I_{m}$, where $1_{m}$ denotes the all-one row vector of length $m$ and $I_{m}$ represents the identity matrix of order $m$. By applying Theorem \ref{lem3.1}, we derive the conclusion.
\end{proof}
\begin{remark}
In \cite{Sobhani2016}, Sobhani showed that $\mathcal{C}_{\itau(t,\tau)}$ is monomially equivalent to the matrix-product code
$(\mathcal{D}_{p^s-1},\ldots,\mathcal{D}_0)\odot CYC(p,s)$,
where
$$\mathcal{D}_{i}=\begin{cases}
	\mathbb{F}_q, & \text{if } i \geq \itau(t,\tau), \\
	\{0\}, & \text{otherwise},
\end{cases}$$
and $CYC(p,s)$ is a $p^s \times p^s$ matrix over $\mathbb{F}_q$ whose $(i,j)$-th entry is
$(-1)^{i+j}\binom{p^s -i}{p^s-j} \bmod p$ for $1 \le i,j \le p^s$.
In Lemma \ref{mat}, we derive two new matrix-product codes that are monomially equivalent to $\mathcal{C}_{\itau(t,\tau)}$. This lemma illustrates that the codewords constituting $\mathcal{C}_{\itau(t,\tau)}$ can be expressed through the codewords of an MDS code. Such a representation achieves improved clarity through the utilization of our matrix-product structure, thereby demonstrating superior performance in the computation of weight distribution.
\end{remark}
Now, we proceed to determine the weight distribution of the code $\mathcal{C}_{\itau(t,\tau)}$.
\begin{theorem}\label{th1}
Let $p$ be a prime and $s\geq 1$ a positive integer. Let $\fq$ denote a finite field of characteristic $p$.
Suppose $1\leq \tau\leq p-1$ and $0\leq t\leq s-1$.
For any non-negative integer $v$, define
\begin{equation}\label{T}
	T_{m,p^t,v}=\left\{(j_1,\cdots,j_{v})\in\mathbb{N}^{v}| j_{1}+\cdots+j_{v}=\frac{m}{p^t}, \,\tau+1 \leq j_{1}, \cdots, j_{v} \leq p \right\},
\end{equation}
 where $T_{m,p^t,0}$ is defined as the empty set. The weight distribution of $\mathcal{C}_{L(t,\tau)}$ is presented in Table~\ref{tab1}.
\end{theorem}
\begin{table}[h]
	\centering
	\caption{Weight distribution of the cyclic code $\mathcal{C}_{L(t,\tau)}$}
	\label{tab1}
	\begin{threeparttable}
		\begin{tabular}{cc}
		\hline
		Weight & Frequency \\
		\hline
		0 & 1 \\
		\makecell{ $(\tau+1)p^t \leq m \leq p^s$ \\ and \\  $m \equiv 0 \pmod{p^t}$} &
			$A_m=\sum_{v=\lfloor\frac{m}{p^{t+1}}\rfloor}^{\lfloor\frac{m}{(\tau+1)p^{t}}\rfloor} \binom{p^{s-t-1}}{v} \sum_{(j_1,\cdots,j_{v})\in T_{m,p^t,v}}\prod_{u=1}^{v} \binom{p}{j_u} \sum_{l=0}^{j_u-\tau-1}(-1)^{l}\binom{j_u}{l}\left(q^{j_u-\tau-l}-1\right).$ \\
		otherwise & $0$\\
		\hline
	   \end{tabular}
		\begin{tablenotes}
			\footnotesize
			\item[] Note that the summation over all $v$-tuples $(j_1,\cdots,j_{v}) \in T_{m,p^t,v}$ evaluates to zero if the set $T_{m,p^t,v}$ is empty. Additionally, the binomial coefficient $\binom{i}{j}$ is formally defined as zero whenever $i < j$.
		\end{tablenotes}			
	\end{threeparttable}
\end{table}
\begin{proof}
Based on Theorem \ref{lem3.1}, the weight distribution of $\mathcal{C}_{L(t,\tau)}$ is identical to that of  $({\hat{\mathcal{C}}}_\tau^{\oplus p^{s-t-1}})^{p^t}$.Therefore, it suffices to investigate the weight distribution of $({\hat{\mathcal{C}}}_\tau^{\oplus p^{s-t-1}})^{p^t}$.
Let $\{A_m|0 \leq m \leq p^{s}\}$ and $\{B_w|0 \leq w \leq p^{s-t}\}$ denote the weight distributions of $\mathcal{C}_{L(t,\tau)}$ and ${\hat{\mathcal{C}}}_\tau^{\oplus p^{s-t-1}}$, respectively. According to the definition of repetition codes, we conclude that
\begin{equation}\label{eq4}
	A_m=\begin{cases}
		B_{\frac{m}{p^t}}, & m \equiv 0 \pmod{p^t} ,\\
		0, & otherwise .
	\end{cases}
\end{equation}
Let $\{D_j|0 \leq j \leq p\}$ be the weight distribution of $\hat{\mathcal{C}}_\tau$.
According to Lemma \ref{lem1}, the following equation holds:
\begin{equation}\label{eq5}
	D_{j}=\begin{cases}
		1, & \text{if } j=0,\\
		0, & \text{if } 0 < j < \tau+1, \\
		\binom{p}{j} \sum_{l=0}^{j-\tau-1}(-1)^{l}\binom{j}{l}\left(q^{j-\tau-l}-1\right), & \text{otherwise.}
	\end{cases}
\end{equation}
According to the relationship between ${\hat{\mathcal{C}}}_\tau^{\oplus p^{s-t-1}}$ and $\hat{\mathcal{C}}_\tau$, it follows that
$$
B_w=\sum_{(j_1,\cdots,j_{p^{s-t-1}})\in S_w}\prod_{u=1}^{p^{s-t-1}}D_{j_u},
$$
where
$$
S_w=\{(j_1,\cdots,j_{p^{s-t-1}})\in\mathbb{Z}^{p^{s-t-1}}|j_1+\cdots+j_{p^{s-t-1}}=w, 0 \leq j_1,\cdots,j_{p^{s-t-1}} \leq p\}.
$$
If $0 < j_u < \tau+1$ for some $1 \leq u \leq p^{s-t-1}$, it follows that $\prod_{u=1}^{p^{s-t-1}}D_{j_u}=0$.  Consequently,
$$
B_w=\sum_{(j_1,\cdots,j_{p^{s-t-1}})\in S_{w}^{'}}\prod_{u=1}^{p^{s-t-1}}D_{j_u},
$$
where
\begin{align*}
	S_{w}^{'}=\{(j_1,\cdots,j_{p^{s-t-1}})\in\mathbb{Z}^{p^{s-t-1}}| & j_1+\cdots+j_{p^{s-t-1}}=w, \text{ and }\\
	& j_u=0 \text{ or } \tau+1 \leq j_u \leq p, \text{for all } 1 \leq u \leq p^{s-t-1}\}.
\end{align*}
When $w=0$, it can be deduced that $S_0^{'}=\{(0,0,\cdots,0)\}$, and consequently $B_0=1$. For the case where $1 \leq w <\tau+1$, it follows that $S_w^{'}=\emptyset$, leading to $B_w=0$. When $\tau+1 \leq w \leq p^{s-t}$, let $v$ denote the number of non-zero components in the vector $(j_1,\cdots,j_{p^{s-t-1}})$. By means of  elementary number-theoretic analysis, it can be concluded that
$$
\left\lfloor\frac{w}{p}\right\rfloor \leq v \leq \left\lfloor\frac{w}{\tau+1}\right\rfloor.
$$
For $\tau+1 \leq w \leq p^{s-t}$, define
\begin{align*}
	S_{w,v}^{'}=\{(j_1,\cdots,j_{p^{s-t-1}})\in\mathbb{Z}^{p^{s-t-1}}|
	& j_{u_1}+\cdots+j_{u_v}=w, \tau+1 \leq j_{u_1}, \cdots, j_{u_v} \leq p, \text{ and } \\
	& j_u=0 \text{ for }  u \in [p^{s-t-1}]\setminus\{u_1,\cdots,u_v\} \}.
\end{align*}
Then $S_{w,v}^{'}\neq\emptyset$ if and only if $\left\lfloor\frac{w}{p}\right\rfloor \leq v \leq \left\lfloor\frac{w}{\tau+1}\right\rfloor$. Consequently,
$$
S_{w}^{'}=\bigcup_{v=\left\lfloor\frac{w}{p}\right\rfloor}^{\left\lfloor\frac{w}{\tau+1}\right\rfloor}S_{w,v}^{'}
$$
and the $\left\lfloor\frac{w}{\tau+1}\right\rfloor-\left\lfloor\frac{w}{p}\right\rfloor$ sets $S_{w,v}^{'}$ are pairwise disjoint. Therefore,
$$
B_w=\sum_{v=\lfloor\frac{w}{p}\rfloor}^{\lfloor\frac{w}{\tau+1}\rfloor} \sum_{(j_1,\cdots,j_{p^{s-t-1}})\in S_{w,v}^{'}}\prod_{u=1}^{p^{s-t-1}}D_{j_u}.
$$
If two vectors
$$(j_1,\cdots,j_v,0,\cdots,0)\in S_{w,v}^{'}$$
and
$$
(0,\cdots,0,j_{u_1},0,\cdots,0,j_{u_2},\cdots\cdots\cdots,j_{u_v},0,\cdots,0)\in S_{w,v}^{'}
$$
satisfy the condition $j_t=j_{u_t}$ for all $1 \leq t \leq v$, it follows that
$$
\prod_{t=1}^{v}D_{j_t}=\prod_{t=1}^{v}D_{j_{u_t}}.
$$
Let
$$
T_{w,v}=\{(j_1,\cdots,j_{v})\in\mathbb{Z}^{v}| j_{1}+\cdots+j_{v}=w, \tau+1 \leq j_{1}, \cdots, j_{v} \leq p \}.
$$
Then
\begin{equation}\label{eq3}
	B_w=\sum_{v=\lfloor\frac{w}{p}\rfloor}^{\lfloor\frac{w}{\tau+1}\rfloor} \binom{p^{s-t-1}}{v} \sum_{(j_1,\cdots,j_{v})\in T_{w,v}}\prod_{u=1}^{v}D_{j_u}.
\end{equation}
By combining Equations \eqref{eq4}, \eqref{eq5}, and \eqref{eq3}, for $(\tau+1)p^t \leq m \leq p^s$ and
$m \equiv 0 \bmod{p^t}$, we obtain
$$
A_m=\sum_{v=\lfloor\frac{m}{p^{t+1}}\rfloor}^{\lfloor\frac{m}{(\tau+1)p^{t}}\rfloor} \binom{p^{s-t-1}}{v} \sum_{(j_1,\cdots,j_{v})\in T_{m,p^t,v}}\prod_{u=1}^{v} \binom{p}{j_u} \sum_{l=0}^{j_u-\tau-1}(-1)^{l}\binom{j_u}{l}\left(q^{j_u-\tau-l}-1\right).
$$
\end{proof}

\begin{remark}\label{rem1}
	The result presented in \cite[Proposition 6.5 (i) and (iii)]{dinh2008linear} can be derived as a direct consequence of Theorem \ref{th1} by setting $\tau = p - 1$. Since case (i) is a special instance of case (iii), we only show that the conclusion given in case (iii) of Proposition 6.5 in \cite{dinh2008linear} follows from Theorem \ref{th1}. For clarity, we restate (iii) using our notation: the code $\mathcal{C}_{L(t,\,p-1),\,s}$ has the following weight distribution:
	
	\begin{equation}\label{am}
		A_{m} =
		\begin{cases}
			\binom{p^{s-t-1}}{h}(q-1)^h, & \text{if } m = p^{t+1}h \text{ for } 0 \leq h \leq p^{s-t-1}, \\
			0, & \text{otherwise}.
		\end{cases}
	\end{equation}
	We demonstrate the derivation of Equation \eqref{am} using the formula provided in Table~\ref{tab1}.
	According to Table~\ref{tab1}, the conditions $p^{t+1} \leq m \leq p^s$ and $m \equiv 0 \pmod{p^t}$ are equivalent to $m = p^{t+1}h$ for $1 \leq h \leq p^{s-t-1}$. The possible value of $v$ satisfying the inequality $\left\lfloor\frac{m}{p^{t+1}}\right\rfloor \leq v \leq \left\lfloor\frac{m}{(\tau+1)p^{t}}\right\rfloor$ is precisely equal to $h$, and the set $T_{p^{t+1}h,p^t,h} = \{(\underbrace{p,\ldots,p}_{h \text{ times}})\}$. Therefore,  
	$$
	A_{p^{t+1}h} = \binom{p^{s-t-1}}{h}(q-1)^h \prod_{u=1}^{h}\binom{p}{p}(-1)^0\binom{p}{0}(q-1), \text{ for } 1 \leq h \leq p^{s-t-1},
	$$  
	which simplifies to $\binom{p^{s-t-1}}{h}(q-1)^h$. Moreover, $A_0 = 1$, and all other cases yield zero, which confirms the Equation \eqref{am}.
\end{remark}

\begin{example}
	Let $p=2$, $q=4$, $s=3$, $\tau=1$ and $t=1$. Consider the cyclic code $\mathcal{C}_{6,3}=\langle(x-1)^6\rangle$ of length $8$ over $\mathbb{F}_4$. According to Theorem \ref{th1}, the potential nonzero weights of $\mathcal{C}$ are $4$, $6$ and $8$. It can be verified that when $m=6$, the corresponding set $T_{m,p^t,v}$ (as defined in \eqref{T}) is empty, leading to the conclusion that $A_6=0$. For $m=4$ and $m=8$, applying the formula for $A_m$ provided in Theorem \ref{th1}, we derive $A_4=6$ and $A_8=9$. Consequently, the weight enumerator of $\mathcal{C}$ is expressed as $1+6z^4+9z^8$. This indicates that $\mathcal{C}_{6,3}$ is a two-weight cyclic code.
	
\end{example}

\begin{example}
	Let $p=3$, $q=9$, $s=3$, $\tau=1$ and $t=1$. Consider the cyclic code $\mathcal{C}=\langle(x-1)^{21}\rangle$ of length $27$ over $\mathbb{F}_9$. According to Theorem \ref{th1}, the potential nonzero weights of $\mathcal{C}$ are $6,9,12,15,18,21,24$ and $27$. By utilizing the formula for $A_m$ presented in Theorem \ref{th1}, we derive that the weight enumerator of $\mathcal{C}$ is given by $1+72z^6+168z^9+1728z^{12}+8064z^{15}+23232z^{18}+96768z^{21}+225792z^{24}+175616z^{27}$.
\end{example}
\begin{theorem}\label{constant-weight}
	Let $s\geq 3$ be a positive integer. 
	The code $\mathcal{C}_{p^s-p,\,s}$ is a $p$-weight cyclic code with parameters $[p^s,p,p^{s-1}]$ and the weight distribution of $\mathcal{C}_{p^s-p,\,s}$ is given by Table \ref{table1}.
\end{theorem}
\begin{table}[h]
	\centering
	\caption{Weight distribution of $\mathcal{C}_{p^s-p,\,s}$}
	\begin{tabular}{ccc}
		\hline
		Weight & Frequency & Notations\\
		\hline
		0 & 1 &\\
		\hline
	$A_{\gamma p^{s-1}}$ & $\binom{p}{\gamma}(q-1)^\gamma$ &  $1\leq \gamma \leq p$\\
		\hline
	\end{tabular}
	\label{table1}
\end{table}
\begin{proof}
	Let $t=s-2$ and $\tau=p-1$. Then $L(t,\tau)=p^s-p$. It is necessary to demonstrate that $\mathcal{C}_{L(s-2,\,p-1)}$ constitutes a $p$-weight cyclic code with the weight distribution presented in Table \ref{table1}. Based on Theorem~\ref{th1}, the frequency $A_m$ of $\mathcal{C}_{p^s-p,\,s}$ is nonzero exclusively when $p^{s-1} \leq m \leq p^s$ and $m \equiv 0 \bmod{p^{s-2}}$. Denote
	$m = p^{s-1} + k p^{s-2}$,
	where $0 \leq k \leq p^2 - p$. Subsequently, both $\lfloor\frac{m}{p^{t+1}}\rfloor$ and $\lfloor\frac{m}{(\tau+1)p^{t}}\rfloor$ are equal to $1 + \lfloor \frac{k}{p}\rfloor$. To proceed, we analyze the value of $k$ under two distinct cases.
		
	\textbf{Case 1.}  If \( k = (\gamma - 1)p \), where \( 1 \leq \gamma \leq p \), then the values \( \lfloor \frac{m}{p^{t+1}} \rfloor \) and \( \lfloor \frac{m}{(\tau + 1)p^t} \rfloor \) are both equal to \( \gamma \). It can be verified that \( T_{m,p^t, v} \) contains exactly one \( v \)-tuple in which all entries are equal to \( p \). Consequently, the frequency \( A_m \) is given by
	\[ A_m = \binom{p}{\gamma} \prod_{u=1}^{\gamma} \binom{p}{p} (-1)^0 \binom{p}{0} (q-1) = \binom{p}{\gamma} (q-1)^{\gamma}. \]
	
	\textbf{Case 2.} If \( (\gamma - 1)p < k \leq \gamma p - 1 \), where \( 1 \leq \gamma \leq p - 1 \), then the value of \( v \) satisfying the inequality \( \lfloor \frac{m}{p^{t+1}} \rfloor  \leq v\leq \lfloor \frac{m}{(\tau + 1)p^t} \rfloor \)
	equals \( \gamma \), and \( T_{m,p^t, \gamma} \) is empty. Therefore, the frequency \( A_m \) is zero in this case.
\end{proof}
We provide illustrative examples of two-weight cyclic codes and three-weight cyclic codes. All these examples have been verified using the Magma \cite{magma97}.
\begin{example}
	Let $\mathcal{C}$ be a cyclic code as defined in Example 2. By applying Theorem \ref{constant-weight}, it follows directly that $\mathcal{C}$ is a two-weight code with the weight enumerator given by $1+6z^4+9z^8$.
\end{example}

\begin{example}
	Let $\mathcal{C}$ be a cyclic code of length $27$ over $\mathbb{F}_9$ with generator polynomial $\langle(x-1)^{24}\rangle$. By applying Theorem \ref{constant-weight}, it follows directly that $\mathcal{C}$ is a three-weight code with the weight enumerator given by $1+24z^9+192z^{18}+512z^{27}$.
\end{example}

For specific values $i = p^s - L(t,\tau)$, which correspond to the dual code of Case I and belong to Case II, an explicit formula for the weight distribution can be derived from Theorem \ref{th1} by utilizing the duality property. We now present the weight distribution of dual codes.

\begin{lemma}[\cite{huffman2010fundamentals}]\label{lem4}
	Given a code $C$ of length $n$, and let $A_i$ and $A_i^{'}$ be the number of codewords of Hamming weight $i$ in $C$ and $C^{\perp}$, respectively, then
	\begin{equation}\label{dual-weight}
		A_k^{'}=\frac{1}{|C|}\sum_{m=0}^{n}A_m P_k(m,n),
	\end{equation}
	where $P_k(x,n)$ is the Krawtchouk polynomial in $x$ of degree $k$, defined by
	\begin{equation}\label{krawtchouk}
		P_k(x,n)=\sum_{j=0}^{k}(-1)^j(q-1)^{k-j}\binom{x}{j} \binom{n-x}{k-j},
	\end{equation}
	where $\binom{i}{j}$ denotes the binomial coefficient, which is assumed to be zero when $i$ is strictly less than $j$.
\end{lemma}

\begin{theorem}\label{th3}
	For integers $1\leq \tau\leq p-1$ and $0\leq t\leq s-1$, let $\mathcal{C}_{(p-\tau) p^{s-t-1}}$ be a cyclic code over $\mathbb{F}_{q}$ of length $p^{s}$. For $p-\tau+1 \leq m \leq p^s$, $\tau+1\leq w\leq p^{s-t}$ and $\lfloor\frac{w}{p}\rfloor \leq v\leq \lfloor\frac{w}{\tau+1}\rfloor$, let
	\begin{align}
		&T_{w,v}=  \left\{(j_1,\cdots,j_{v})\in\mathbb{Z}^{v}| j_{1}+\cdots+j_{v}=w, \tau+1 \leq j_{1}, \cdots, j_{v} \leq p \right\}, \label{a}\\
		&B_w=  \sum_{v=\lfloor\frac{w}{p}\rfloor}^{\lfloor\frac{w}{\tau+1}\rfloor} \binom{p^{s-t-1}}{v} \sum_{(j_1,\cdots,j_{v})\in T_{w,v}}\prod_{u=1}^{v} \binom{p}{j_u} \sum_{l=0}^{j_u-\tau-1}(-1)^{l}\binom{j_u}{l}\left(q^{j_u-\tau-l}-1\right), \label{b}\\
		&P_m(wp^t,p^s)=  \sum_{l=0}^{m}(-1)^l(q-1)^{m-l}\binom{wp^t}{l} \binom{p^s-wp^t}{m-l}. \label{c}
	\end{align}
	The weight distribution of $\mathcal{C}_{(p-\tau) p^{s-t-1}}$ is given by Table \ref{tab_case2}.
\end{theorem}

\begin{table}[!ht]
	\centering
	\caption{Weight Distribution of $\mathcal{C}_{(p-\tau) p^{s-t-1}}$}
	\begin{tabular}{c|c}
		\hline
		Weight & Frequency \\
		\hline
		0 & 1 \\
		$1$  & 0 \\
		$2 \leq m \leq p^s$ &$A_m=\frac{1}{q^{(p-\tau) p^{s-t-1}}}\left((q-1)^m\binom{p^s}{m}+\sum_{w=\tau+1}^{p^{s-t}}B_w P_m(wp^t,p^s)\right)$\\
		\hline
	\end{tabular}
	\label{tab_case2}
\end{table}

\begin{proof}
	The minimum distance of $\mathcal{C}_{(p-\tau) p^{s-t-1}}$ is $p - \tau + 1\geq 2$ when $t = 0$, and it equals $2$ for all values of $t$ in the range $1 \leq t \leq s - 1$. Consequently, $A_1 = 0$ for all $t$ satisfying $0 \leq t \leq s - 1$. For $2 \leq m \leq p^s$, the dual code of $\mathcal{C}_{(p-\tau) p^{s-t-1}}$ is $\mathcal{C}_{L(t,\tau)}$. Let $A_k^{'}$ be the number of codewords with Hamming weight $k$ in $\mathcal{C}_{L(t,\tau)}$. Based on Theorem \ref{th1} and  Lemma \ref{lem4}, it follows that
	\begin{align*}
		A_m&=\frac{1}{q^{(p-\tau) p^{s-t-1}}}\sum_{k=0}^{p^s}A_k^{'}P_m(k,p^s)\\
		&=\frac{1}{q^{(p-\tau) p^{s-t-1}}}\left((q-1)^m\binom{p^s}{m}+\sum_{w=\tau+1}^{p^{s-t}}A_{wp^t}^{'}P_m(wp^t,p^s)\right)\\
		&=\frac{1}{q^{(p-\tau) p^{s-t-1}}}\left((q-1)^m\binom{p^s}{m}+\sum_{w=\tau+1}^{p^{s-t}}B_w P_m(wp^t,p^s)\right),
	\end{align*}
	where $B_w$ and $P_m(wp^t,p^s)$ are defined as in \eqref{b} and \eqref{c}, respectively.
\end{proof}

\begin{remark}\label{rem2}
	The result presented in \cite[Proposition 6.5 (ii) and (v)]{dinh2008linear} can be derived as a direct consequence of Theorem \ref{th3} by setting $\tau = p - 1$. The corresponding analysis is similar to that outlined in Remark \ref{rem1}.
\end{remark}

\begin{example}
	Let $\tau=p-1$, $s=2$ and $t=1$. Consider the cyclic code $\mathcal{C}=\langle x-1\rangle $ of length $p^2$ over finite field $\mathbb{F}_q$, where $q$ is a power of the prime $p$.
	According to Theorem \ref{th3}, the weight distribution of $\mathcal{C}$ is obtained and presented in the following table.
	\begin{table}[h]
		\centering
		\begin{tabular}{cc}
			\hline
			Weight & Frequency \\
			\hline
			0 & 1 \\
			1 & 0 \\
			$2 \leq m \leq p^2$ &$A_m=\frac{1}{q}\binom{p^2}{m} [(q-1)^m+(-1)^m(q-1)]$\\
			\hline
		\end{tabular}
		\label{tab2}
	\end{table}
	
	\noindent Furthermore, assuming that $p=3$ and $q=9$, the weight enumerator of $\mathcal{C}$ is $1+288z^2+4704z^3+57456z^4+458640z^5+2446752z^6+8388576z^7+16777224z^8+14913080z^9$. The result has been verified using the Magma \cite{magma97}.
\end{example}
\section{Weight Distribution for Case II}
In this section, we examine the weight distribution of $\mathcal{C}_i$ for the range $\itau(t,\tau-1) < i < \itau(t,\tau)$, where $0 \leq t \leq s - 1$ and $1 \leq \tau \leq p - 1$. Let $i=\itau(t,\tau-1)+i^{'}$, where $0 < i' < p^{s-t-1}$.
For a fixed $t \in [0, s - 2]$, we commence with the analysis of a particular case in which $\tau = 1$ and $i' = (p - \varsigma)p^{s - t' - 1}$, where $t + 1 \leq t' \leq s - 1$ and $1 \leq \varsigma \leq p - 1$. This case falls within the interval $\itau(t,0) < i < \itau(t,1)$ for $t \in [0, s - 2]$.
\begin{theorem}\label{rest}
For integers $1\leq \varsigma\leq p-1$, $0\leq t\leq s-2$, and $t+1 \leq t' \leq s-1$, let $\mathcal{C}_{p^s-p^{s-t}+(p-\varsigma)p^{s-t'-1}}$ be a cyclic code over $\mathbb{F}_{q}$ of length $p^{s}$. For 
$0\leq m\leq p^s$,
$2 \leq m^{'} \leq p^{s-t}$, $\varsigma+1 \leq w \leq p^{s-t'}$ and $\lfloor\frac{w}{p}\rfloor \leq v \leq \lfloor\frac{w}{\varsigma+1}\rfloor$, let 
\begin{align*}
	&T_{w,v}=\{(j_1,\cdots,j_{v})\in\mathbb{Z}^{v}| j_{1}+\cdots+j_{v}=w, \varsigma+1 \leq j_{1}, \cdots, j_{v} \leq p \},\\
	&D_w=\sum_{v=\lfloor\frac{w}{p}\rfloor}^{\lfloor\frac{w}{\varsigma+1}\rfloor} \binom{p^{s-t'-1}}{v} \sum_{(j_1,\cdots,j_{v})\in T_{w,v}}\prod_{u=1}^{v} \binom{p}{j_u} \sum_{l=0}^{j_u-\varsigma-1}(-1)^{l}\binom{j_u}{l}\left(q^{j_u-\varsigma-l}-1\right),\\
	&P_{m^{'}}(w p^{t'-t},p^{s-t})=\sum_{l=0}^{m^{'}}(-1)^l(q-1)^{m^{'}-l}\binom{w p^{t'-t}}{l} \binom{p^{s-t}-w p^{t'-t}}{m^{'}-l}.
\end{align*}
The weight distribution of $\mathcal{C}_{p^s-p^{s-t}+(p-\varsigma)p^{s-t'-1}}$ is given by Table \ref{tab4}.	
	\begin{table}[h]
		\centering
		\caption{Weight distribution of the cyclic code $\mathcal{C}_{p^s-p^{s-t}+(p-\varsigma)p^{s-t'-1}}$}
		\begin{tabular}{cc}
			\toprule
			Weight & Frequency \\
			\midrule
			0 & 1 \\
			\makecell{ $2p^t \leq m \leq p^s$ and \\  $m \equiv 0 \pmod{p^t}$}
			&$A_m=\frac{1}{q^{(p-\varsigma)p^{s-t'-1}}}\left((q-1)^{
				m^{'}}\binom{p^{s-t}}{m^{'}}+\sum_{w=\varsigma+1}^{p^{s-t'}}D_w P_{
				m^{'}}(w p^{t'-t},\,p^{s-t})\right)$, where $m^{'}=\frac{m}{p^t}$\\  
			otherwise & 0\\
			\bottomrule
		\end{tabular}
		\label{tab4}
	\end{table}	
\end{theorem}

\begin{proof}
	According to Theorem \ref{d}, the cyclic code $\mathcal{C}_{p^s-p^{s-t}+(p-\varsigma)p^{s-t'-1}}$ is monomially equivalent to $\mathcal{D}^{p^t}$, where $\mathcal{D}=\langle(x-1)^{(p-\varsigma)p^{s-t'-1}}\rangle$ is a cyclic code of length $p^{s-t}$ over $\mathbb{F}_{q}$. 
	Let $\{B_{m^{'}}|0 \leq m^{'} \leq p^{s-t}\}$ be the weight distribution of $\mathcal{D}$. Based on Theorem \ref{th1}, the weight distribution of $\mathcal{C}_{p^s-p^{s-t}+(p-\varsigma)p^{s-t'-1}}$ is as follows:
	\begin{equation}\label{eq1}
		A_m=\begin{cases}
			B_{\frac{m}{p^t}}, & m \equiv 0 \pmod{p^t} ,\\
			0, & otherwise .
		\end{cases}
	\end{equation}
	According to Theorem \ref{th3}, the weight distribution of $\mathcal{D}$ is listed as follows: $B_0=1$, $B_1=0$, and for $2 \leq m^{'} \leq p^{s-t}$, 
		\begin{equation}\label{eq2}
			B_{m^{'}}=\frac{1}{q^{(p-\varsigma)p^{s-t'-1}}}\left((q-1)^{m^{'}}\binom{p^{s-t}}{m^{'}}+\sum_{w=\varsigma+1}^{p^{s-t'}}D_w P_{m^{'}}(w p^{t'-t},p^{s-t})\right),
		\end{equation}
		where
		\begin{align*}
			&D_w=\sum_{v=\lfloor\frac{w}{p}\rfloor}^{\lfloor\frac{w}{\varsigma+1}\rfloor} \binom{p^{s-t'-1}}{v} \sum_{(j_1,\cdots,j_{v})\in T_{w,v}}\prod_{u=1}^{v} \binom{p}{j_u} \sum_{l=0}^{j_u-\varsigma-1}(-1)^{l}\binom{j_u}{l}\left(q^{j_u-\varsigma-l}-1\right),\\
			& T_{w,v}=\{(j_1,\cdots,j_{v})\in\mathbb{Z}^{v}| j_{1}+\cdots+j_{v}=w, \varsigma+1 \leq j_{1}, \cdots, j_{v} \leq p \},\\
			&P_{m^{'}}(w p^{t'-t},\,p^{s-t})=\sum_{l=0}^{m^{'}}(-1)^l(q-1)^{m^{'}-l}\binom{w p^{t'-t}}{l} \binom{p^{s-t}-w p^{t'-t}}{m^{'}-l}.
		\end{align*}
	By combining equations \eqref{eq1} and \eqref{eq2}, for $2p^t \leq m \leq p^s$ and $m \equiv 0 \pmod{p^t}$, we define $m' = \frac{m}{p^t}$ and obtain that
	$
	A_m=\frac{1}{q^{(p-\varsigma)p^{s-t'-1}}}\left((q-1)^{
		m^{'}}\binom{p^{s-t}}{m^{'}}+\sum_{w=\varsigma+1}^{p^{s-t'}}D_w P_{
		m^{'}}(w p^{t'-t},p^{s-t})\right).
	$
	\end{proof}

When $\varsigma=p-1$ and $t^{'}=s-1$, \cite[Proposition 6.5 (iv)]{dinh2008linear} can be derived directly from Theorem \ref{th1} and Theorem~\ref{rest}.

\begin{corollary}\cite[Proposition 6.5 (iv)]{dinh2008linear}\label{cor1}
	For an integer $0\leq t\leq s-1$, let $\mathcal{C}_{p^s-p^{s-t}+1}$ be a cyclic code over $\mathbb{F}_{q}$ of length $p^{s}$. The weight distribution of $\mathcal{C}_{p^s-p^{s-t}+1}$ is given by Table \ref{tab5}.	
\end{corollary}	
	\begin{table}[!ht]
	\centering
	\caption{Weight distribution of the cyclic code $\mathcal{C}_{p^s-p^{s-t}+1}$}
	\begin{tabular}{cc}
		\toprule[0.5pt]
		Weight & Frequency \\
		\midrule[0.3pt]
		0 & 1 \\
		\makecell{ $2p^t \leq m \leq p^s$, and \\  $m \equiv 0 \pmod{p^t}$}
		&$A_m=\frac{1}{q}\binom{p^{s-t}}{m^{'}}\left((q-1)^{m^{'}}+(-1)^{m^{'}}(q-1)\right)$, where $m^{'}=\frac{m}{p^t}$\\
		otherwise & 0\\
		\bottomrule[0.5pt]
	\end{tabular}
	\label{tab5}
\end{table}
\begin{proof}
	The analysis of weight distribution of $\mathcal{C}_{p^s-p^{s-t}+1}$ is divided into two cases: $t=s-1$ and $0\leq t\leq s-2$.
	
	When $t=s-1$, the code $\mathcal{C}_{p^s-p+1}$ is $\mathcal{C}_{L(s-1,1)}$. According to Theorem \ref{th1},  when $2p^{s-1} \leq m \leq p^s$ and $m \equiv 0 \pmod{p^{s-1}}$, denote $m^{'}=\frac{m}{p^t}$, we obtain
	\begin{align}
		A_m&=\sum_{v=\lfloor\frac{m}{p^{t+1}}\rfloor}^{\lfloor\frac{m}{(\tau+1)p^{t}}\rfloor} \binom{1}{v} \sum_{(j_1,\cdots,j_{v})\in T_{m,p^t,v}}\prod_{u=1}^{v} \binom{p}{j_u} \sum_{l=0}^{j_u-\tau-1}(-1)^{l}\binom{j_u}{l}\left(q^{j_u-\tau-l}-1\right)\\
		&=\binom{p}{m^{'}} \sum_{l=0}^{m^{'}-2}(-1)^{l}\binom{m^{'}}{l}\left(q^{m^{'}-1-l}-1\right), \label{17}
	\end{align}
	where \eqref{17} follows from the fact that $\binom{a}{b}=0$ if $a<b$ and $T_{m,p^t,1}=\{(m^{'})\}$.
	It follows from $(1+(-1))^{m^{'}}=0$ that $ \sum_{l=0}^{m^{'}-2}(-1)^{l}\binom{m^{'}}{l}(-1)=(-1)^{m^{'}-1}(m^{'}-1)$.
	Therefore, the equation \eqref{17} can be transformed into
	\begin{align}
		A_m&=
		\frac{1}{q}\binom{p}{m^{'}} \left( \sum_{l=0}^{m^{'}-2}(-1)^{l}\binom{m^{'}}{l}q^{m^{'}-l}+(-1)^{m^{'}-1}(m^{'}-1)q \right)\\
		&=\frac{1}{q}\binom{p}{m^{'}} \left( \sum_{l=0}^{m^{'}}(-1)^{l}\binom{m^{'}}{l}q^{m^{'}-l} +(-1)^{m^{'}}q-(-1)^{m^{'}}\right)\\
		&=\frac{1}{q}\binom{p}{m^{'}}\left((q-1)^{m^{'}}+(-1)^{m^{'}}(q-1)\right)\label{22}.
	\end{align}
    
    When $0\leq t\leq s-2$, according to Theorem \ref{rest},
	set $\varsigma=p-1$ and $t^{'}=s-1$, the expression for $A_m$ when $m$ lies in the range $2p^t \leq m \leq p^s$ and satisfies the congruence condition $m \equiv 0 \pmod{p^t}$ is given by  
	$$
	\frac{1}{q}\left((q - 1)^{m^{'}} \binom{p^{s - t}}{m^{'}} + D_p P_{m^{'}}(p^{s - t}, p^{s - t})\right).
	$$
	Furthermore, we can verify that $D_p = q - 1$ since $T_{p,1} = \{(j_1) = (p)\}$, and that $P_{m^{'}}(p^{s - t}, p^{s - t}) = (-1)^{m^{'}} \binom{p^{s - t}}{m^{'}}$, noting that $\binom{a}{b} = 0$ when $a < b$. Consequently, we obtain  
	\begin{equation}\label{am6}
		A_m = \frac{1}{q} \binom{p^{s - t}}{m^{'}} \left( (q - 1)^{m^{'}} + (-1)^{m^{'}} (q - 1) \right).
	\end{equation}
	By combining \eqref{22} and \eqref{am6}, we derive the expression of $A_m$ across all values of $t$ within the range $0 \leq t \leq s-1$, as summarized in Table~\ref{tab5}.
\end{proof}

By utilizing the dual properties of the weight distribution and applying Theorem \ref{rest}, we derive the weight distribution of $\mathcal{C}_i$ for the case $i = p^{s-t} - p^{s-t'} + \varsigma p^{s-t'-1}$, where $0 \leq t \leq s - 2$, $t + 1 \leq t' \leq s - 1$, and $1 \leq \varsigma \leq p - 1$. This particular case falls within the interval $\itau(0,0) < i < \itau(0,1)$.
\begin{theorem}\label{dual}
	For integers $1\leq \varsigma\leq p-1$, $0\leq t\leq s-2$, and $t+1 \leq t' \leq s-1$, let $\mathcal{C}_{p^{s-t}-p^{s-t'}+\varsigma p^{s-t'-1}}$ be a cyclic code over $\mathbb{F}_{q}$ of length $p^{s}$. For $2 \leq m^{'} \leq p^{s-t}$, $\varsigma+1 \leq w \leq p^{s-t'}$ and $\lfloor\frac{w}{p}\rfloor \leq v \leq \lfloor\frac{w}{\varsigma+1}\rfloor$, let 
	\begin{align*}
		&T_{w,v}=\{(j_1,\cdots,j_{v})\in\mathbb{Z}^{v}| j_{1}+\cdots+j_{v}=w, \varsigma+1 \leq j_{1}, \cdots, j_{v} \leq p \},\\
		&D_w=\sum_{v=\lfloor\frac{w}{p}\rfloor}^{\lfloor\frac{w}{\varsigma+1}\rfloor} \binom{p^{s-t'-1}}{v} \sum_{(j_1,\cdots,j_{v})\in T_{w,v}}\prod_{u=1}^{v} \binom{p}{j_u} \sum_{l=0}^{j_u-\varsigma-1}(-1)^{l}\binom{j_u}{l}\left(q^{j_u-\varsigma-l}-1\right),\\
		&P_k(x,n)=\sum_{l=0}^{k}(-1)^l(q-1)^{k-l}\binom{x}{l} \binom{n-x}{k-l}.
	\end{align*}
	The weight distribution of $\mathcal{C}_{p^s-p^{s-t}+(p-\varsigma)p^{s-t'-1}}$ is given by Table \ref{tab3}.	
	\begin{table}[h]
		\centering
		\caption{Weight distribution of $\mathcal{C}_{p^{s-t}-p^{s-t'}+\varsigma p^{s-t'-1}}$}
		\begin{tabular}{cc}
			\toprule[0.5pt]
			Weight & Frequency \\
			\hline
			0 & 1 \\
			1 & 0 \\
			$2 \leq m \leq p^s$
			& \makecell{$A_m=\frac{1}{q^{p^{s-t}}}\sum_{m'=2}^{p^{s-t}}\left((q-1)^{m'}\binom{p^{s-t}}{m'}+\sum_{w=\varsigma+1}^{p^{s-t'}}D_w P_{m'}(w p^{t'-t},p^{s-t})\right)P_{m}(m' p^{t},p^{s})$ \\ $+\frac{1}{q^{p^{s-t}-p^{s-t'}+\varsigma p^{s-t'-1}}}(q-1)^{m}\binom{p^{s}}{m}$}
			\\
			\bottomrule[0.5pt]
		\end{tabular}
		\label{tab3}
	\end{table}
\end{theorem}
\begin{proof}
	For $0\leq k\leq p^s$, let $A_k^{'}$ denote the number of codewords with Hamming weight $k$ in the dual code of the code $\mathcal{C}_{p^{s-t}-p^{s-t'}+\varsigma p^{s-t'-1}}$, which corresponds to the code $\mathcal{C}_{p^s-p^{s-t}+(p-\varsigma)p^{s-t'-1}}$.
	According to Lemma \ref{lem4} and Theorem \ref{rest}, the following holds:
	\begin{align}
			A_m=&\, \frac{1}{q^{p^{s-t}-p^{s-t'}+\varsigma p^{s-t'-1}}}\sum_{k=0}^{p^s}A_k^{'}P_{m}(k,p^s) \label{e1}\\
		=&\, \frac{1}{q^{p^{s-t}-p^{s-t'}+\varsigma p^{s-t'-1}}}\left(P_{m}(0,p^s)+\sum_{m'=2}^{p^{s-t}}A_{m'p^t}^{'}P_{m}(m'p^t,p^s) \right) \label{e2}\\
		=& \,   \frac{1}{q^{p^{s-t}}}\sum_{m'=2}^{p^{s-t}}\left((q-1)^{m'}\binom{p^{s-t}}{m'}+\sum_{w=\varsigma+1}^{p^{s-t'}}D_w P_{m'}(w p^{t'-t},p^{s-t})\right)P_{m}(m' p^{t},p^{s}) \nonumber\\
		&  +\frac{1}{q^{p^{s-t}-p^{s-t'}+\varsigma p^{s-t'-1}}}(q-1)^{m}\binom{p^{s}}{m}, \label{e3}
	\end{align}
	where \eqref{e1} is derived from the dual property presented in equation \eqref{dual-weight}, \eqref{e2} is based on the fact that $A_1^{'} = 0$, and \eqref{e3} follows from the expression for $A_m$ given in Table~\ref{tab4}.
\end{proof}

When $\varsigma=p-1$ and $t^{'}=s-1$, \cite[Proposition 6.5 (vi)]{dinh2008linear} can be derived directly from Theorem \ref{th3} and Theorem~\ref{dual}.
\begin{corollary}\cite[Proposition 6.5 (vi)]{dinh2008linear}\label{cor2}
	For an integer $0\leq t\leq s-1$, let $\mathcal{C}_{p^{s-t}-1}$ be a cyclic code over $\mathbb{F}_{q}$ of length $p^{s}$. The weight distribution of $\mathcal{C}_{p^{s-t}-1}$ is given by Table \ref{tab4}.	
	\begin{table}[h]
		\centering
		\caption{Weight distribution of $\mathcal{C}_{p^{s-t}-1}$}
		\begin{tabular}{cc}
			\hline
			Weight & Frequency \\
			\hline
			0 & 1 \\
			1 & 0 \\
			$2 \leq m \leq p^s$
			&$A_m=\frac{1}{q^{p^{s-t}}}\sum_{m^{'}=0}^{p^{s-t}}\binom{p^{s-t}}{m^{'}}\left((q-1)^{m^{'}}+(-1)^{m^{'}}(q-1)\right)P_m(m^{'}p^t,p^s)$\\
			\hline
		\end{tabular}
		\label{tab7}
	\end{table}
\end{corollary}	
\begin{proof}
	The analysis of weight distribution of $\mathcal{C}_{p^{s-t}-1}$ is divided into two cases: $t=s-1$ and $0\leq t\leq s-2$.
	  
	  When $t=s-1$, the weight distribution of $\mathcal{C}_{p-1}$ can be derived using 
	Theorem~\ref{th3} by setting $\tau=1$ and $t=s-1$. According to Table~\ref{tab_case2}, we obtain that $A_0=1$, $A_1=0$ and for $m\geq 2$, 
	\begin{equation}\label{am4}
		A_m = \frac{1}{q^{p-1}} \left( (q-1)^m \binom{p^s}{m} + \sum_{w=2}^{p} B_w P_m (w p^{s-1}, p^s) \right) ,
	\end{equation}
	where 
	$B_w= \binom{p}{w} \sum_{l=0}^{w-2} (-1)^l \binom{w}{l} (q^{w-1-l} - 1)$. 
	Verify that 
	\begin{equation}\label{am5}
		\frac{1}{q^{p-1}} (q-1)^m \binom{p^s}{m}= \frac{1}{q^{p}} \sum_{w=0}^{1} \binom{p}{w} \left( (q-1)^{w} + (-1)^{w} (q-1) \right) P_m( w p^{s-1} ,\, p^{s}),
	\end{equation}
	By combining equations \eqref{am4} and \eqref{am5}, we obtain  
	\begin{align}
		A_m & = \frac{1}{q^{p}} \sum_{w=0}^{1} \binom{p}{w} \left( (q-1)^{w} + (-1)^{w} (q-1) \right) P_m( w p^{s-1} ,\, p^{s})  +\\
		& \indent \frac{q}{q^{p}} \sum_{w=2}^{p} \binom{p}{w}
		\left( \sum_{l=0}^{w-2} (-1)^l \binom{w}{l} (q^{w-1-l} - 1) \right) P_m (w p^{s-1},\, p^s). \label{am1}
	\end{align}
	Consider the expression of \eqref{am1}, it can be reformulated as follows:
	\begin{align}
		\eqref{am1}&=
		\frac{1}{q^{p}} \sum_{w=2}^{p} \binom{p}{w}\left( \sum_{l=0}^{w-2} (-1)^l \binom{w}{l} q^{w-l} - q\sum_{l=0}^{w-2} (-1)^l \binom{w}{l}) \right) P_m (w p^{s-1}, p^s)\\
		&= \frac{1}{q^{p}} \sum_{w=2}^{p} \binom{p}{w}
		\left( \sum_{l=0}^{w-2} (-1)^l \binom{w}{l} q^{w-l} +(-1)^{w-1} wq + (-1)^{w}q \right) P_m (w p^{s-1}, p^s)\label{am3}\\
        &= \frac{1}{q^{p}} \sum_{w=2}^{p} \binom{p}{w}
        \left( \sum_{l=0}^{w} (-1)^l \binom{w}{l} q^{w-l} + (-1)^{w} (q-1)
         \right) P_m (w p^{s-1}, p^s)\\
        & =\frac{1}{q^{p}} \sum_{w=2}^{p} \binom{p}{w}\left( (q-1)^{w} + (-1)^{w} (q-1) \right) P_m( w p^{s-1} , p^{s}),
	\end{align}
	where \eqref{am3} follows from the identity $0=(1-1)^w=\sum_{l=0}^{w-2} (-1)^{l} \binom{w}{l} + (-1)^{w-1} w + (-1)^{w}$.
	Therefore, 
	\begin{equation*}
		A_m=\frac{1}{q^{p}} \sum_{w=0}^{p} \binom{p}{w} \left((q-1)^{w} + (-1)^{w} (q-1)\right) P_m( w p^{s-1} ,\, p^{s}).
	\end{equation*}
	
	When $0\leq t\leq s-2$, according to Theorem \ref{dual}, set $\varsigma = p - 1$ and $t' = s - 1$, the expression for $A_m$ in the case $2 \leq m \leq p^s$ is given by  
	\begin{equation}\label{Am}
		A_m=\frac{1}{q^{p^{s-t}-1}}(q-1)^m\binom{p^s}{m} + \frac{1}{q^{p^{s-t}}}\sum_{m'=2}^{p^{s-t}}\binom{p^{s-t}}{m'}\left((q-1)^{m'} + (-1)^{m'}(q-1)\right)P_m(m'p^t, p^s).
	\end{equation}
	Consider the expression of the summation of Equation \eqref{Am}, which is 
	\[\frac{1}{q^{p^{s-t}}}\sum_{m'=2}^{p^{s-t}}\binom{p^{s-t}}{m'}\left((q-1)^{m'} + (-1)^{m'}(q-1)\right)P_m(m'p^t, p^s).\] It can be verified that this expression evaluates to zero when $m' = 1$. When $m' = 0$, the expression coincides with the term $\frac{1}{q^{p^{s-t}-1}}(q-1)^m\binom{p^s}{m}$. This observation allows Equation \eqref{Am} to be reformulated in a unified form, 
    \begin{equation}
    	A_m=\frac{1}{q^{p^{s-t}}}\sum_{m'=0}^{p^{s-t}}\binom{p^{s-t}}{m'}\left((q-1)^{m'} + (-1)^{m'}(q-1)\right)P_m(m'p^t, p^s).
    \end{equation}
	Then we obtain the formula of $A_m$, as shown in Table \ref{tab7}.
	\end{proof}

In the general case $\itau(t, \tau - 1) < i < \itau(t, \tau)$ for $0\leq t\leq s-1$ and $1\leq \tau\leq p-1$, the situation becomes more intricate. We proceed to investigate a more detailed structure of codewords of $\mathcal{C}_i$ in order to determine the weight distribution of $\mathcal{C}_i$ for $i$ within the range $\itau(t, \tau - 1) < i < \itau(t, \tau)$.

\begin{theorem}\label{ciProperty}
Let $0\leq t\leq s-2$, $1\leq \tau\leq p-1$ and $\itau(t,\tau-1)<i<\itau(t,\tau)$. The cyclic code $\mathcal{C}_i=\langle (x-1)^i\rangle \subseteq \fq[x]/\langle x^{p^s}-1 \rangle$ of length $p^s$ over $\mathbb{F}_q$ is monomially equivalent to the linear code $\bar{\mathcal{D}}^{p^t}$, where $\bar{\mathcal{D}}$ is defined by the following properties:
\begin{itemize}
	\item [(i)] $\hat{\mathcal{C}}_{\tau}^{\oplus p^{s-t-1}}\subsetneqq\bar{\mathcal{D}}\subsetneqq\hat{\mathcal{C}}_{\tau-1}^{\oplus p^{s-t-1}}$;
	\item[(ii)] For all $0\leq l\leq p^{s-t-1}-1$, $\bar{\mathcal{D}}|_{S_{l}}=\widehat{\mathcal{C}}_{\tau-1}$, where $S_l=\{lp+j\,|\,0\leq j\leq p -1, j\in \mathbb{Z} \}$.
\end{itemize}
\end{theorem}

\begin{proof}
Let $i=p^s-p^{s-t}+(\tau-1)p^{s-t-1}+i'$, where $0 < i' < p^{s-t-1}$. 
The first result (i) follows from the fact that 
\begin{equation}\label{c1}
	\mathcal{C}_{\itau(t,\tau)}\subsetneqq \mathcal{C}_i\subsetneqq \mathcal{C}_{\itau(t,\tau-1)}.
\end{equation}

We now proceed to prove (ii).
According to Theorem \ref{d},
$\mathcal{C}_{i}$ is monomially equivalent to $\mathcal{D}^{p^t}$, where $\mathcal{D}\!=\!\langle(x\!-\!1)^{(\tau\!-\!1)p^{s\!-\!t\!-\!1}\!+i'}\!\rangle$ is a cyclic code of length $p^{s-t}$ over $\mathbb{F}_{q}$. Denote $\gamma=p^{s-t-1}$. Let 
\begin{equation}\label{gx1}
	g(x)=(x-1)^{(\tau-1)\gamma+i'}f(x)
\end{equation}
be a codeword of $\mathcal{D}$, where $f(x)=\sum_{\theta=0}^{(p-\tau+1)\gamma-i'-1}f_{\theta} x^{\theta}\in\mathbb{F}_q[x]$.
Let $f_l(x)$ be a sum of the monomial terms of polynomial $f(x)$ whose degree modulo $\gamma$ is equal to $l$, where $0 \le l \le \gamma-1$.
Then
$$f_l(x)= \begin{cases}
	\displaystyle{\sum_{\lambda=0}^{p-\tau}f_{\lambda\gamma+l} x^{\lambda\gamma+l} },& \text{  if } 0 \le l \le \gamma-1-i' ,\\
	\displaystyle{\sum_{\lambda=0}^{p-\tau-1}f_{\lambda\gamma+l} x^{\lambda\gamma+l} },& \text{  if }\gamma-i' \le l \le \gamma-1 .
\end{cases}
$$
Let
$$\bar{f_l}(x)= \begin{cases}
	\displaystyle{\sum_{\lambda=0}^{p-\tau}f_{\lambda\gamma+l} x^{\lambda}}, & \text{  if } 0 \le l \le \gamma-1-i' ,\\
	\displaystyle{\sum_{\lambda=0}^{p-\tau-1}f_{\lambda\gamma+l} x^{\lambda} }, & \text{  if } \gamma-i' \le l \le \gamma-1 .
\end{cases}
$$
Then, the codeword $g(x)$ (defined in \eqref{gx1}) of $\mathcal{D}$ can be expressed in the following form: 
\begin{align}\label{gx}
	g(x)=(x^{\gamma}-1)^{\tau-1} \left(\sum_{\theta=0}^{i'}\binom{i'}{\theta}(-1)^{i'-\theta}x^{\theta}\right) \left(\sum_{l=0}^{\gamma-1}\bar{f_l}(x^{\gamma})x^l\right).
\end{align}
Denote $a_{\theta}=\binom{i'}{\theta}(-1)^{i'-\theta}$. The expansion of the product $(\sum_{\theta=0}^{i'}a_{\theta}x^{\theta})\times(\sum_{l=0}^{\gamma-1}\bar{f_l}(x^{\gamma})x^l)$ yields a total of $(i^{'}+1)\times\gamma$ terms. To facilitate further analysis, we need to rearrange these terms in accordance with Table~\ref{ta}.
\begin{table}[t!]
	\centering
	\caption{}
	\label{ta}
	\begin{tabular}{c|ccccc}
		\Xhline{1pt}
		$\theta+l\equiv 0 (\bmod \gamma)$ & $\theta=0,l=0$ & $\theta=1,l=\gamma-1$ & $\theta=2,l=\gamma-2$ & $\cdots$ & $\theta=i^{'},l=\gamma-i^{'}$\\
		\Xhline{0.5pt}
		$\theta+l\equiv 1 (\bmod \gamma)$ & $\theta=0,l=1$ & $\theta=1,l=0$ & $\theta=2,l=\gamma-1$ & $\cdots$ & $\theta=i^{'},l=\gamma-i^{'}+1$\\
		\Xhline{0.5pt}
		$\theta+l\equiv 2 (\bmod \gamma)$ & $\theta=0,l=2$ & $\theta=1,l=1$ & $\theta=2,l=0$ & $\cdots$ & $\theta=i^{'},l=\gamma-i^{'}+2$\\
		\Xhline{0.5pt}
		$\vdots$ & $\vdots$ & $\vdots$ & $\vdots$ & $\vdots$ & $\vdots$ \\
		\Xhline{0.5pt}
		$\theta+l\equiv \gamma-2 (\bmod \gamma)$ & $\theta=0,l=\gamma-2$ & $\theta=1,l=\gamma-3$ & $\theta=2,l=\gamma-4$ & $\cdots$ & $\theta=i^{'},l=\gamma-i^{'}-2$\\
		\Xhline{0.5pt}
		$\theta+l\equiv \gamma-1 (\bmod \gamma)$ & $\theta=0,l=\gamma-1$ & $\theta=1,l=\gamma-2$ & $\theta=2,l=\gamma-3$ & $\cdots$ & $\theta=i^{'},l=\gamma-i^{'}-1$\\
		\Xhline{1pt}
	\end{tabular}
\end{table}
Thus,
\begin{align}
	g(x)=&\left[a_0\bar{f_0}(x^{\gamma}) + a_1\bar{f}_{\gamma-1}(x^{\gamma})x^{\gamma}+\cdots + a_{i'}\bar{f}_{\gamma-i'}(x^{\gamma})x^{\gamma}\right](x^{\gamma}-1)^{\tau-1} \label{g1}\\
	& \qquad \vdots \qquad\qquad\qquad \vdots \qquad\qquad\qquad\qquad \vdots \\
	+&  \left[a_0\bar{f}_{i'-1}(x^{\gamma}) + a_1\bar{f}_{i'-2}(x^\gamma)+\cdots + a_{i'}\bar{f}_{\gamma-1}(x^{\gamma})x^{\gamma}\right](x^{\gamma}-1)^{\tau-1}x^{i'-1}\\
	+&  \left[a_0\bar{f}_{i'}(x^{\gamma}) + a_1\bar{f}_{i'-1}(x^\gamma)+\cdots + a_{i'}\bar{f}_{0}(x^{\gamma})\right](x^{\gamma}-1)^{\tau-1}x^{i'}\\
	& \qquad \vdots \qquad\qquad\qquad \vdots \qquad\qquad\qquad\qquad \vdots \\
	+&  \left[a_0\bar{f}_{\gamma-1}(x^{\gamma}) + a_1\bar{f}_{\gamma-2}(x^{\gamma}) + \cdots + a_{i'}\bar{f}_{\gamma-1-i'}(x^{\gamma})\right](x^{\gamma}-1)^{\tau-1}x^{\gamma-1} \label{g2}.
\end{align}
Namely,
\begin{align*}
	g(x)=&\sum_{l=0}^{i'-1}\left[\sum_{\theta=0}^{l}a_{\theta}\bar{f}_{l-\theta}(x^{\gamma})+\sum_{\theta=l+1}^{i'}a_{\theta}\bar{f}_{\gamma+l-\theta}(x^{\gamma})x^{\gamma}\right](x^{\gamma}-1)^{\tau-1}x^{l} \\
	&+\sum_{l=i'}^{\gamma-1}\left[\sum_{\theta=0}^{i'}a_{\theta}\bar{f}_{l-\theta}(x^{\gamma})\right](x^{\gamma}-1)^{\tau-1}x^{l}.
\end{align*}
Let $\bar{\mathcal{D}}$ be a linear code monomially equivalent to $\mathcal{D}$ with the corresponding monomial matrix denoted as
$P=(\delta_{\alpha \beta})_{p^{s-t} \times p^{s-t}}$, where
$$\delta_{\alpha \beta}=\left\{
\begin{array}{ll}
	1, & \text{if } \alpha= l p + j \text{ and } \beta = j \gamma +l,  \\
	& \text{    where } 0 \le j \le p-1 \text{ and } 0 \le l \le p^{s-t-1} -1, \\
	0, & \text{otherwise.}
\end{array}\right.$$
Let $\bar{g}(x)$ be the codeword in $\bar{\mathcal{D}}$ which is monomially equivalent to $g(x)$ in $\mathcal{D}$. Then,
\begin{align}
	\bar{g}(x)=&\sum_{l=0}^{i'-1}\left[\sum_{\theta=0}^{l}a_{\theta}\bar{f}_{l-\theta}(x)+\sum_{\theta=l+1}^{i'}a_{\theta}\bar{f}_{\gamma+l-\theta}(x)x\right](x-1)^{\tau-1}x^{lp} \\
	&+\sum_{l=i'}^{\gamma-1}\left[\sum_{\theta=0}^{i'}a_{\theta}\bar{f}_{l-\theta}(x)\right](x-1)^{\tau-1}x^{lp}.
\end{align}
We consider two distinct cases: $0 \leq l \leq i' - 1$ and $i' \leq l \leq \gamma - 1$.
\begin{itemize}
	\item For $0\leq l \leq i'-1$, the punctured code $\bar{\mathcal{D}|}_{S_l}$ consists of codewords in the form $q(x)(x-1)^{\tau-1}$, where
	$$q(x)=\sum_{\theta=0}^{l}\bar{f}_{\theta}(x)+\sum_{\theta=l+1}^{i'}\bar{f}_{\gamma+l-\theta}(x)x.$$
	When $f(x)$ ranges over all polynomials of degree less than $(p-\tau+1)\rho-i'$, the corresponding polynomial $\bar{f_0}(x)$ similarly encompasses all polynomials of degree less than $p-\tau+1$. Consequently, this ensures that $q(x)$ runs through all polynomials of degree less than $p-\tau+1$. Note that the degree of $q(x)$ is less than $p-\tau+1$, then
	\begin{equation}\label{c3}
		\bar{\mathcal{D}|}_{S_l}=\left\{q(x)(x-1)^{\tau-1}|\,q(x)\in \mathbb{F}_q[x] \text{ and  deg}(q(x))<p-\tau+1\right\}=\hat{\mathcal{C}}_{\tau-1}.
	\end{equation}
	\item For $i' \leq l \leq \gamma-1$, the punctured code $\bar{\mathcal{D}|}_{S_l}$ consists of codewords in the form $\sum_{\theta=0}^{i'}\bar{f}_{l-\theta}(x)(x-1)^{\tau-1}$. When $f(x)$ runs through all polynomials of degree less than $(p-\tau+1)\rho-i'$, the corresponding polynomial $\bar{f}_{l-i'}(x)$ similarly encompasses all polynomials of degree less than $p-\tau+1$. Consequently, this ensures that  $\sum_{\theta=0}^{i'}\bar{f}_{l-\theta}(x)$ runs through all polynomials of degree less than $p-\tau+1$. Note that the degree of $\sum_{\theta=0}^{i'}\bar{f}_{l-\theta}(x)$ is less than $p-\tau+1$, then
	\begin{equation}\label{c4}
		\bar{\mathcal{D}|}_{S_l}=\left\{q(x)(x-1)^{\tau-1}|\,q(x)\in \mathbb{F}_q[x] \text{ and } \text{deg}(q(x))<p-\tau+1\right\}=\hat{\mathcal{C}}_{\tau-1}.
	\end{equation}
\end{itemize}

The conclusion stated in (ii) is directly derived from (\ref{c3}) and (\ref{c4}).
\end{proof}

\begin{remark}
	Based on Theorem \ref{ciProperty}, the code $\bar{\mathcal{D}}$ comprises $p^{s-t-1}$ concatenated codewords of $\widehat{\mathcal{C}}_{\tau-1}$. Furthermore, each codeword in $\bar{\mathcal{D}}$ can be expressed as $(\mathbf{c}_0,\mathbf{c}_1,\ldots,\mathbf{c}_{p^{s-t-1}-1})$, where $\mathbf{c}_j \in \widehat{\mathcal{C}}_{\tau-1}$ for all $0 \leq j \leq p^{s-t-1}-1$. 
	Additionally, these $p^{s-t-1}$ vectors $\mathbf{c}_j$, where $0 \leq j \leq p^{s-t-1}-1$, are interrelated. More precisely, let $X = (\xi_0, \xi_1, \ldots, \xi_{p^{s-t-1}-1})$ denote a random codeword sampled from $\bar{\mathcal{D}}$. Since $\bar{\mathcal{D}}$ is a proper subset of $\hat{\mathcal{C}}_{\tau-1}^{\oplus p^{s-t-1}}$, it follows from equations \eqref{g1}-\eqref{g2} that the variables $\xi_0, \xi_1, \ldots, \xi_{p^{s-t-1}-1}$ are mutually dependent. 	
\end{remark}

For the two subsets $\bar{\mathcal{D}}$ and $\hat{\mathcal{C}}_{\tau}^{\oplus p^{s-t-1}}$ of $\mathbb{F}_q[x]/\langle x^{p^{s-t}}-1\rangle$, let their difference be denoted by $I$. Specifically,
\[ I = \{(x-1)^{(\tau-1)p^{s-t-1}+\xi} f(x) \mid f(x) \in \fq[x], \deg(f(x)) < (p-\tau+1)p^{s-t-1}-\xi, \text{ and } (x-1)^{p^{s-t-1}-\xi} \nmid f(x)\}. \]
Recall that $\bar{\mathcal{D}} = \langle(x-1)^{(\tau-1)p^{s-t-1}+\xi}\rangle$ is a cyclic code of length $p^{s-t}$, where $0 < \xi < p^{s-t-1}$. Furthermore, $\bar{\mathcal{D}}$ contains the subcode $\hat{\mathcal{C}}_{\tau}^{\oplus p^{s-t-1}}$, whose weight distribution is characterized in Theorem~\ref{th1}. 
Consequently, the determination of the weight distribution of $\bar{\mathcal{D}}$ can be reduced to the evaluation of the weight distribution of the smaller set $I$. It should be noted that the Hamming weight of a polynomial is defined as the number of its nonzero coefficients.

\begin{theorem}\label{th2}
	Let $t,\tau,\xi$ be integers satisfying $0\leq t\leq s-2$, $1\leq \tau\leq p-1$, and $1 \leq \xi < p^{s-t-1}$. Let $\mathcal{C}_{L(t,\tau-1)+\xi}=\langle (x-1)^{L(t,\tau-1)+\xi} \rangle \subseteq \mathbb{F}_q[x]/\langle x^{p^s}-1 \rangle$ be a cyclic code of length $p^s$ over $\mathbb{F}_q$. For any $1\leq m\leq p^s$,
	let $v$ be a non-negative integer satisfying $\lfloor\frac{m}{p^{t+1}}\rfloor \leq v \leq \lfloor\frac{m}{(\tau+1)p^t}\rfloor$. We define the set
	\begin{equation*}
		T_{m,p^t,v}=\left\{(j_1,\cdots,j_{v})|\, \tau+1 \leq j_{1}, \cdots, j_{v} \leq p, \text{ and } j_{1}+\cdots+j_{v}=\frac{m}{p^t} \right\},
	\end{equation*}
	where $T_{\frac{m}{p^t},0}$ is explicitly defined as the empty set.
	 Let
	\begin{equation*}
		I = \{(x-1)^{(\tau-1)p^{s-t-1}+\xi} f(x) \mid f(x) \in \mathbb{F}_q[x], \deg(f(x)) < (p-\tau+1)p^{s-t-1}-\xi, \text{ and } (x-1)^{p^{s-t-1}-\xi} \nmid f(x)\}.
	\end{equation*}
	Let $D_{\frac{m}{p^t}}$ denote the number of polynomials in $I$ with Hamming weight $\frac{m}{p^t}$.
	The weight distribution of $\mathcal{C}_{L(t,\tau-1)+\xi}$ is given by Table \ref{table_L}.
	
	\begin{table}[h]
		\centering
		\caption{Weight Distribution of the Cyclic Code $\mathcal{C}_{L(t,\tau-1)+\xi}$}
		\begin{threeparttable}
			\begin{tabular}{c|c}
				\hline
				Weight & Frequency \\
				\hline
				0 & 1 \\
				\makecell{$m \equiv 0 \pmod{p^t}$ and \\ $(\tau+1)p^t \leq m \leq p^s$} &
				\makecell{$A_m=\sum_{v=\lfloor\frac{m}{p^{t+1}}\rfloor}^{\lfloor\frac{m}{(\tau+1)p^{t}}\rfloor} \binom{p^{s-t-1}}{v} \sum_{(j_1,\cdots,j_{v})\in T_{m,p^t,v}}\prod_{u=1}^{v} \binom{p}{j_u} \sum_{l=0}^{j_u-\tau-1}(-1)^{l}\binom{j_u}{l}\left(q^{j_u-\tau-l}-1\right)+D_{\frac{m}{p^t}}$} \\
				\hline
			\end{tabular}
		\end{threeparttable}
		\label{table_L}
	\end{table}
	
\end{theorem}
\begin{proof}
	Based on Theorem \ref{ciProperty}, $\mathcal{C}_{L(t,\tau-1)+\xi}$ is monomially equivalent to $\bar{\mathcal{D}}^{p^t}$. Hence the weight distribution of $\mathcal{C}_{L(t,\tau-1)+\xi}$ is equal to that of $\bar{\mathcal{D}}^{p^t}$. Let $\{B_w | 0 \leq w \leq p^{s-t}\}$ be the weight distribution $\bar{\mathcal{D}}$. According to the definition of repetition codes,
	\[A_m=\begin{cases}
		B_{\frac{m}{p^t}}, & m \equiv 0 \pmod{p^t} ,\\
		0, & otherwise .
	\end{cases}\]
	The code $\bar{\mathcal{D}}$ contains the subcode $\hat{\mathcal{C}}_{\tau}^{\oplus p^{s-t-1}}$ by Theorem \ref{ciProperty}. The number of codewords with weight $\frac{m}{p^t}$ in $\hat{\mathcal{C}}_{\tau}^{\oplus p^{s-t-1}}$ is
	\[\sum_{v=\lfloor\frac{m}{p^{t+1}}\rfloor}^{\lfloor\frac{m}{(\tau+1)p^t}\rfloor} \binom{p^{s-t-1}}{v} \sum_{(j_1,\cdots,j_{v})\in T_{m,p^t,v}}\prod_{u=1}^{v} \binom{p}{j_u} \sum_{l=0}^{j_u-\tau-1}(-1)^{l}\binom{j_u}{l}\left(q^{j_u-\tau-l}-1\right),\]
	where
	\[T_{m,p^t,v}=\{(j_1,\cdots,j_{v})\in\mathbb{Z}^{v}| j_{1}+\cdots+j_{v}=\frac{m}{p^t}, \tau+1 \leq j_{1}, \cdots, j_{v} \leq p \}.\]
	The weight distribution of the difference set $I$ obtained by subtracting the subcode $\hat{\mathcal{C}}_{\tau}^{\oplus p^{s-t-1}}$ from the code $\bar{\mathcal{D}}$ is $\{D_{w} | 0 \leq w \leq p^{s-t}\}$. Hence, the number of codewords with weight $\frac{m}{p^t}$ in $\bar{\mathcal{D}}$ is
	\[A_m=\sum_{v=\lfloor\frac{m}{p^{t+1}}\rfloor}^{\lfloor\frac{m}{(\tau+1)p^t}\rfloor} \binom{p^{s-t-1}}{v} \sum_{(j_1,\cdots,j_{v})\in T_{m,p^t,v}}\prod_{u=1}^{v} \binom{p}{j_u} \sum_{l=0}^{j_u-\tau-1}(-1)^{l}\binom{j_u}{l}\left(q^{j_u-\tau-l}-1\right) + D_{\frac{m}{p^t}}.\]
\end{proof}

The weight distribution of $\mathcal{C}_{L(t,\tau-1)+\xi}$ is provided in Table \ref{table_L}. As shown in Table \ref{table_L}, the expression $A_m$ comprises two distinct components. The first component corresponds to the weight distribution of $\hat{\mathcal{C}}_{\tau}^{\oplus p^{s-t-1}}$, which can be derived from Table \ref{tab1}. The second component corresponds to the weight distribution of the polynomial set $I$, defined as follows:  
\begin{align}
	I = \Big\{  (x-1)^{(\tau-1)p^{s-t-1}+\xi} & f(x) \,\Big| \, f(x) \in \mathbb{F}_q[x],   \nonumber\\
	&\deg(f(x)) < (p-\tau+1)p^{s-t-1}-\xi, \text{ and } (x-1)^{p^{s-t-1}-\xi} \nmid f(x) \Big\}. \label{I1}
\end{align}
To facilitate a more systematic analysis of all elements in $I$, we proceed to reformulate its representation, aiming to simplify the computation of its weight distribution.

The polynomial in the set $I$ can be expressed as $(x-1)^{(\tau-1)p^{s-t-1}+\xi} f(x) \in I$, where $f(x) = f_0 + f_1 x + \cdots + f_{\theta} x^{\theta} \in \mathbb{F}_q[x]$ and $\theta = (p - \tau + 1)p^{s-t-1} - \xi - 1$. The expression of $f(x)$ can be reformulated as follows:
\begin{equation}\label{fx}
	f(x) = f_0^{'} + f_1^{'} (x - 1) + \cdots + f_{\theta}^{'} (x - 1)^{\theta} \in \mathbb{F}_q[x],
\end{equation}
where
$$
\begin{pmatrix}
	f_0^{'} \\
	f_1^{'} \\
	\vdots \\
	f_{\theta}^{'}
\end{pmatrix}
=
\begin{pmatrix}
	1 & \binom{1}{0}(-1)^1 & \cdots & \binom{\theta}{0}(-1)^{\theta} \\
	0 & 1 & \cdots & \binom{\theta}{1}(-1)^{\theta - 1} \\
	\vdots & \vdots & \ddots & \vdots \\
	0 & 0 & \cdots & 1
\end{pmatrix}^{-1}
\begin{pmatrix}
	f_0 \\
	f_1 \\
	\vdots \\
	f_{\theta}
\end{pmatrix}.
$$
The polynomial $(x-1)^{(\tau-1)p^{s-t-1}+\xi} f(x)$ belongs to $I$ if and only if $(x-1)^{p^{s-t-1}-\xi}$ does not divide $f(x)$. Based on the specific expression of $f(x)$ given in \eqref{fx}, this non-divisibility condition is equivalent to the requirement that not all of the coefficients $f_0', f_1', \dots, f_{p^{s-t-1}-\xi-1}'$ are simultaneously zero. Consequently, the set $I$ can be redefined as follows:
\begin{align}
	I = \Big\{  f_0'(x-1)^{(\tau-1)p^{s-t-1}+\xi} + \cdots + &f_{(p-\tau+1)p^{s-t-1}-\xi-1}'(x-1)^{p^{s-t}-1} \in \mathbb{F}_q[x] \,\Big| \nonumber\\
	& f_0', f_1', \dots, f_{p^{s-t-1}-\xi-1}' \text{ are not all zero} \Big\}. \label{I2}
\end{align}

By employing the set $I$ defined in equation \eqref{I2}, the computation of the weight distribution of $\mathcal{C}_{L(t,\tau-1)+\xi}$ becomes more efficient. In order to further clarify the specific steps involved in determining the weight distribution of the cyclic code $\mathcal{C}_{L(t,\tau-1)+\xi}$, the following illustrative example is provided.
\begin{example}\label{example_S}
	 Set $t=1,\tau=2,\xi=1$. Consider the cyclic code $C_{22}=\langle(x-1)^{22}\rangle$ of length $27$ over the finite filed $\mathbb{F}_3$. Denote the weight distribution of $\mathcal{C}_{22}$ as $(A_0,A_1,\ldots,A_{27})$.	 
	 According to Theorem \ref{th2}, $A_0=1$ and $A_m=0$ for all $m$ such that $m \not\equiv 0 \pmod{3}$ or $1 \leq m < 9$. 
	 For $9 \leq m \leq 27$ and $m \equiv 0 \pmod{3}$, verify that 
	\[T_{\frac{m}{3},v}=\begin{cases}
		\emptyset, & \text{if } m \not\equiv 0 \pmod{9}, \\
		\{(\underbrace{3,\cdots,3}_{v \text{ times}})\}, & \text{if } m \equiv 0 \pmod{9} \text{ and } v=\frac{m}{9}.
	\end{cases}\]
	Hence
	\begin{equation}\label{am2}
		A_m= \binom{3}{m/9} 2^{\frac{m}{9}} + B_{\frac{m}{3}},
	\end{equation}
	where $B_{\frac{m}{3}}$ is the number of polynomials with Hamming weight $\frac{m}{3}$ in the set
	\[I=\{f_0 (x-1)^{22}+f_1(x-1)^{23}+\cdots+f_4(x-1)^{26} \mid f_0,f_1,\cdots,f_{4} \in \fq, \text{ and } f_0,f_1 \text{ are not both zero} \}.\]
	\begin{table}[h!]
		\centering
		\caption{}
		\begin{tabular}{c|ccccccc}
			\hline
			$m$ &9& 12 & 15 & 18 & 21 &24&27 \\
			$\frac{m}{3}$ &3& 4 & 5 & 6 & 7 &8&9 \\
			$B_{\frac{m}{3}}$ & 0&54 & 0 & 54 & 108 &0&0\\
			\hline
		\end{tabular}
		\label{V}
	\end{table}
	Using the Magma program, we computed $B_{\frac{m}{3}}$ for $9 \leq m \leq 27$ and $m \equiv 0 \pmod{3}$ in the Table \ref{V}.
	Based on \eqref{am2} and the value of $B_{\frac{m}{3}}$ presented in Table \ref{V}, the weight distribution of the cyclic code $C_{22}$ is summarized in Table~\ref{VI}.
	\begin{table}[h!]
		\centering
		\caption{}
		\begin{tabular}{c|ccccccc}
			\hline
			$m$ &  0 & 9 & 12 & 18 & 21 & 27 & otherwise\\
			$A_m$ & 1 & 6 & 54 & 66 & 108 & 8 & 0\\
			\hline
		\end{tabular}
		\label{VI}
	\end{table}
\end{example}

\section{Conclusion}
In this paper, we provide a comprehensive analysis of the structure and weight distribution for repeated-root cyclic codes with prime power lengths. Additionally, we propose a class of $p$-weight cyclic codes for any prime $p$. To further elucidate our approach, several illustrative examples are provided to demonstrate the computation of the weight distribution for these cyclic codes using our proposed formulation. All results presented in the examples have been  verified using Magma \cite{magma97}. Exploring the computation of weight distributions for repeated-root cyclic codes with general lengths as well as simple-root cyclic codes remains an intriguing area for further investigation.

\bibliographystyle{IEEEtran}
\bibliography{IEEEabrv,ref}

\end{document}